\documentclass[aps,preprint]{revtex4}%
\usepackage{amsfonts}
\usepackage{amsmath}
\usepackage{amssymb}
\usepackage{graphicx}%
\begin{document}
\title{Coherent and semiclassical states of a charged particle in electromagnetic fields}
\author{A.\ S.\ Pereira}
\email{albertoufcg@hotmail.com}
\affiliation{Department of Higher Mathematics and Mathematical Physics, Institute of
Physics and Technology, National Research Tomsk Polytechnic University, Tomsk, Russia.}
\pacs{PACS number}

\begin{abstract}
In the present article we extend our study (BJP \textbf{45} (2015) 369) of
generalized coherent states (GCS) of a one-dimensional particle considering
such important physical system as a 3-dimensional charged particle in electric
and magnetic fields. Constructing GCS in many-dimensional case, we meet
nontrivial technical complications that make the consideration nontrivial and
instructive. The GCS of a system under consideration are constructed. We study
properties of these GCS such as completeness relations, minimization of
uncertainty relations and so on. We point out which family of the obtained GCS
of a charged particle in magnetic field is related with the CS constructed
first by Malkin and Man'ko. We obtain conditions under which some of the GCS
can be considered as semiclassical states.

\end{abstract}
\maketitle

\section{Introduction}

Coherent states (CS) play an important role in modern quantum theory as states
that provide a natural relation between quantum mechanical and classical
descriptions. They have a number of useful properties and as a consequence a
wide range of applications, e.g., in semiclassical description of quantum
systems, in quantization theory, in condensed matter physics, in quantum
computations, and so on, see e.g.,
\cite{Kla1968,Gil1973,Kla1985,Per1986,Ali2000,Nie2000,Gaz2009}. Starting by
the works \cite{Mal1979,Dod1987} CS are defined as eigenvectors of some
annihilation operators that are at the same time integrals of motion. Of
course such defined CS have to satisfy the corresponding Schr\"{o}dinger
equation. In the frame of such a definition one can in principle construct CS
for a general quadratic system. In the article \cite{Bag2015} we, following
these scheme, constructed different families of CS (GCS) for one-dimensional
systems with general time-dependent quadratic Hamiltonian, see too
\cite{Bag2013,Bag2014}. In the article \cite{Bag2014}, we have demonstrated
that the GCS of a free particle can be treated as quantum states that describe
a semiclassical motion, whereas there exist GCS that describe pure quantum motion.

In the present article we extend our study beyond one-dimensional systems,
considering such important physical system as a 3-dimensional charged particle
in electric and magnetic fields. Considering many-dimensional systems we meet
nontrivial technical complications that makes the consideration nontrivial and
instructive. We discuss properties of the constructed GCS such as completeness
relations, minimization of uncertainty relations and so on. As in the
one-dimensional case, we succeeded to find conditions that allow one to
attribute the GCS either to the class of semiclassical quantum states or to
purely quantum states.

This article is organized as follows. In Section 2, we outlook classical and
quantum descriptions of a charged particle in parallel electric and magnetic
fields and obtain for such a system integrals of motion which are creation and
annihilation operators. In section 3, we construct GCS that satisfies the
Schr\"{o}dinger equation and calculate the mean values, standard deviations
and uncertainty relations, as well as we discuss properties of the constructed
GCS such as completeness relations, minimization of uncertainty relations, and
so on. We also demonstrated that the GCS of a charged particle in magnetic
field is related with the CS constructed first by Malkin and Man'ko. In
section 4, we obtain conditions under the electric and magnetic field such
that the GCS can be considered as semiclassical states or purely quantum states.

\section{Charged particle in parallel electric and magnetic fields}

\subsection{Classical and quantum equations of motion}

Consider a charged particle$^{1}$\footnotetext[1]{Throughout the text, $e$
denotes the algebraic electric charge ($e=-\left\vert e\right\vert $ for an
electron), bold letters represent vectors, e. g., $\mathbf{r}=\left(
x,y,z\right)  $, latin indices are $j,k,...=1,2,3$ and greek indices are
$\beta,\eta,...=1,2$.}, with total charge $e$, moving in a $3$-dimensional
Euclidean space,%
\begin{equation}
H=\frac{\mathbf{P}^{2}}{2m}+eA_{0},\ \ \mathbf{P}=\mathbf{p}-\frac{e}%
{c}\mathbf{A}, \label{1.1}%
\end{equation}
interacting with constant, uniform and parallel electric $\mathbf{E}$ and
magnetic $\mathbf{B}$ fields%
\begin{align}
&  \mathbf{E}=\left(  0,0,E\right)  ,\ \mathbf{B}=\left(  0,0,B\right)
,\nonumber\\
&  A_{0}=-zE\sin^{2}\alpha,\ \ \mathbf{A}=\frac{1}{2}\left(  -By,Bx,-2ctE\cos
^{2}\alpha\right)  ,\ \ \alpha\in\left[  0,\pi/2\right]  . \label{1.2}%
\end{align}
The corresponding Hamiltonian can be written in the following form%
\begin{align}
&  H=H_{xy}+H_{z},\ \ H_{xy}=\frac{\mathbf{p}_{\bot}^{2}}{2m}+\frac
{m\omega^{2}\mathbf{r}_{\bot}^{2}}{8}-\frac{\omega}{2}L,\nonumber\\
&  H_{z}=\frac{p_{z}^{2}}{2m}-m\xi z\sin^{2}\alpha+\xi p_{z}t\cos^{2}%
\alpha+\frac{m\xi^{2}}{2}t^{2}\cos^{4}\alpha,\nonumber\\
&  L=xp_{y}-yp_{x},\ \ \omega=\frac{eB}{mc},\text{ \ }\xi=\frac{eE}{m}.
\label{1.3}%
\end{align}
Here $\mathbf{p}$ is the canonical momenta conjugated to the coordinates
$\mathbf{r}$, $\omega$ is the cyclotron frequency, and $\xi$ denotes the
acceleration along the $z$-axis. The velocity $\mathbf{v}$ is related to the
canonical variables as $m\mathbf{v=P}$.

The division of $H$ into two separate parts, $H_{xy}$ and $H_{z}$, indicates
implicitly the independence between particle's motion on the $xy$-plane
(hereafter referred as $xy$-motion) from the motion along the $z$-direction
(hereafter referred as $z$-motion). This fact is explicitly confirmed by the
structure of canonical equations of motion$^{2}$\footnotetext[2]{Quantities
associated with the $xy$-motion are labelled with the symbol \textquotedblleft%
$\perp$\textquotedblright\ (e. g., $\mathbf{r}_{\perp}=\left(  x,y\right)  $)
while quantities associated with the $z$-motion are labelled with the symbol
\textquotedblleft$\parallel$\textquotedblright\ (e. g., $\mathbf{r}%
_{\parallel}=z$).}%
\begin{align}
&  v_{x}\left(  t\right)  =\dot{x}\left(  t\right)  =\frac{\partial
H}{\partial p_{x}}=\frac{p_{x}}{m}+\frac{\omega}{2}y,\text{ \ }\dot{p}%
_{x}\left(  t\right)  =-\frac{\partial H}{\partial x}=\frac{\omega}{2}%
p_{y}-\frac{m\omega^{2}}{4}x,\nonumber\\
&  v_{y}\left(  t\right)  =\dot{y}\left(  t\right)  =\frac{\partial
H}{\partial p_{y}}=\frac{p_{y}}{m}-\frac{\omega}{2}x,\text{ \ }\dot{p}%
_{y}\left(  t\right)  =-\frac{\partial H}{\partial y}=-\frac{\omega}{2}%
p_{x}-\frac{m\omega^{2}}{4}y,\nonumber\\
&  v_{z}\left(  t\right)  =\dot{z}\left(  t\right)  =\frac{\partial
H}{\partial p_{z}}=\frac{p_{z}}{m}+\xi t\cos^{2}\alpha,\text{ \ }\dot{p}%
_{z}\left(  t\right)  =-\frac{\partial H}{\partial z}=m\xi\sin^{2}\alpha.
\label{1.4}%
\end{align}

The general solution for the $z$-motion follows from the third line of Eq.
(\ref{1.4}),%
\begin{align}
&  z\left(  t\right)  =z_{0}+v_{z}^{0}t+\frac{1}{2}\xi t^{2},\text{ \ }%
p_{z}\left(  t\right)  =m\left(  v_{z}^{0}+\xi t\sin^{2}\alpha\right)
,\nonumber\\
&  z_{0}=z\left(  0\right)  ,\ \ v_{z}^{0}=v_{z}\left(  0\right)  =\frac
{p_{z}^{0}}{m},\ \ p_{z}^{0}=p_{z}\left(  0\right)  , \label{1.5}%
\end{align}
corresponds to an accelerated motion along the $z$-direction. As for the
$xy$-motion, it can be presented in two equivalent ways. The first one is
given in terms of the initial Cauchy data $\left(  \mathbf{r}_{\perp}\left(
0\right)  =\mathbf{r}_{0\perp}\text{,}\ \mathbf{v}_{\perp}\left(  0\right)
=\mathbf{v}_{\perp}^{0}\right)  $,%
\begin{align}
&  x\left(  t\right)  =x_{0}+v_{y}^{0}\frac{1-\cos\left(  \omega t\right)
}{\omega}+v_{x}^{0}\frac{\sin\left(  \omega t\right)  }{\omega},\text{
\ }\frac{p_{x}\left(  t\right)  }{m}=\frac{v_{x}^{0}\cos\left(  \omega
t\right)  +v_{y}^{0}\sin\left(  \omega t\right)  +v_{x}^{0}-\omega y_{0}}%
{2},\nonumber\\
&  y\left(  t\right)  =y_{0}-v_{x}^{0}\frac{1-\cos\left(  \omega t\right)
}{\omega}+v_{y}^{0}\frac{\sin\left(  \omega t\right)  }{\omega},\text{
\ }\frac{p_{y}\left(  t\right)  }{m}=\frac{v_{y}^{0}\cos\left(  \omega
t\right)  -v_{x}^{0}\sin\left(  \omega t\right)  +v_{y}^{0}+\omega x_{0}}{2},
\label{1.6}%
\end{align}
which can be equivalently given in terms of the perpendicular velocity
components,%
\begin{align}
&  x\left(  t\right)  =x_{0}-\frac{v_{y}\left(  t\right)  -v_{y}^{0}}{\omega
},\ \ y\left(  t\right)  =y_{0}+\frac{v_{x}\left(  t\right)  -v_{x}^{0}%
}{\omega},\nonumber\\
&  p_{x}\left(  t\right)  =m\frac{v_{x}\left(  t\right)  +v_{x}^{0}-\omega
y_{0}}{2},\ \ p_{y}\left(  t\right)  =m\frac{v_{y}\left(  t\right)  +v_{y}%
^{0}+\omega x_{0}}{2},\nonumber\\
&  v_{x}\left(  t\right)  =v_{y}^{0}\sin\left(  \omega t\right)  +v_{x}%
^{0}\cos\left(  \omega t\right)  ,\ v_{y}\left(  t\right)  =v_{y}^{0}%
\cos\left(  \omega t\right)  -v_{x}^{0}\sin\left(  \omega t\right)  .
\label{1.8a}%
\end{align}

\begin{figure}[!htb]
\centering
\includegraphics{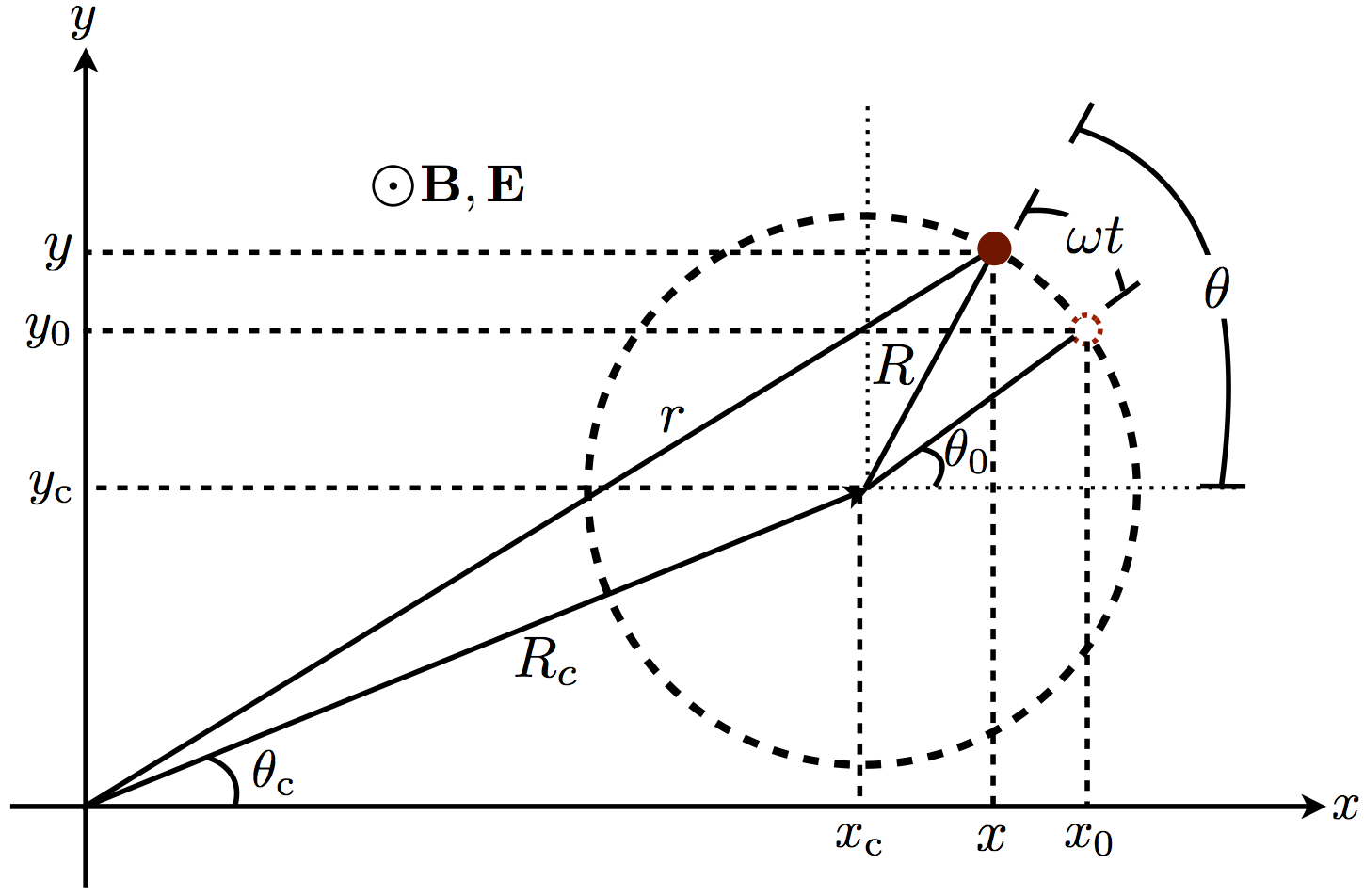}
\caption{Particle's classical circular motion along the $xy$-plane. The $3D$
trajectory correspond to a helix, with a circular $xy$-motion and accelerated
$z$-motion, whose direction point out of the picture, aligned with the
external electromagnetic fields.}
\label{fig1}
\end{figure}

In a second representation, the $xy$-motion is parametrized in terms of the
coordinates of the center of the orbit $\left(  x_{c},y_{c}\right)  $, its
radius $R$ and the initial phase $\theta_{0}$ instead the initial Cauchy data,%
\begin{align}
&  x\left(  t\right)  =x_{c}+R\cos\theta,\ \ y\left(  t\right)  =y_{c}%
+R\sin\theta,\ \ \theta=\omega t+\theta_{0},\nonumber\\
&  x_{c}=R_{c}\cos\theta_{c},\ \ y_{c}=R_{c}\sin\theta_{c},\ \ R_{c}^{2}%
=x_{c}^{2}+y_{c}^{2},\ \ R^{2}=\left(  x-x_{c}\right)  ^{2}+\left(
y-y_{c}\right)  ^{2}, \label{1.9}%
\end{align}
as illustrated in Fig. \ref{fig1}.

In this representation, the distance from the origin $r\left(  t\right)  $ can
be expressed in terms of $R$, $R_{c}$ and the relative angle $\theta
-\theta_{c}$ as%
\begin{equation}
r^{2}=x^{2}+y^{2}=R^{2}+R_{c}^{2}+2RR_{c}\cos\left(  \theta-\theta_{c}\right)
, \label{1.10}%
\end{equation}
whose maximum $r_{\max}=R+R_{c}$ and minimum $r_{\min}=\left\vert
R-R_{c}\right\vert $ correspond to configurations in which $\theta=\theta_{c}$
and $\theta=\theta_{c}+\pi$, respectively. From the perpendicular velocities
$\mathbf{v}_{\perp}\left(  t\right)  =\left(  \dot{x}\left(  t\right)
,\dot{y}\left(  t\right)  \right)  $,%
\begin{align}
\dot{x}\left(  t\right)   &  =v_{x}\left(  t\right)  =-R\omega\sin
\theta=-\omega\left(  y-y_{c}\right)  ,\nonumber\\
\dot{y}\left(  t\right)   &  =v_{y}\left(  t\right)  =R\omega\cos\theta
=\omega\left(  x-x_{c}\right)  , \label{1.11}%
\end{align}
one can readily see that both sets of solutions can be mapped to each other
through the following relations between the initial conditions $\left(
x_{0},y_{0},v_{x}^{0},v_{y}^{0}\right)  $ and the integration constants
$\left(  x_{c},y_{c},\theta_{0},R\right)  $ as%
\begin{align}
&  x_{0}=x_{c}+R\cos\theta_{0},\ \ y_{0}=y_{c}+R\sin\theta_{0},\nonumber\\
&  v_{x}^{0}=-R\omega\sin\theta_{0},\ \ v_{y}^{0}=R\omega\cos\theta_{0}.
\label{1.12}%
\end{align}

Eqs. (\ref{1.9}) and (\ref{1.11}) may be written as follows%
\begin{align}
&  x\left(  t\right)  =x_{c}+\frac{1}{\omega}\sqrt{\frac{2\mathcal{E}_{\perp}%
}{m}}\cos\theta,\ \ y\left(  t\right)  =y_{c}+\frac{1}{\omega}\sqrt
{\frac{2\mathcal{E}_{\perp}}{m}}\sin\theta,\nonumber\\
&  v_{x}\left(  t\right)  =-\sqrt{\frac{2\mathcal{E}_{\perp}}{m}}\sin
\theta,\ \ v_{y}\left(  t\right)  =\sqrt{\frac{2\mathcal{E}_{\perp}}{m}}%
\cos\theta, \label{1.13}%
\end{align}
where the perpendicular energy $\mathcal{E}_{\perp}$ is a conserved quantity,%
\begin{equation}
\mathcal{E}_{\perp}=\frac{m\left(  v_{x}^{2}+v_{y}^{2}\right)  }{2}%
=\frac{m\left[  \left(  v_{x}^{0}\right)  ^{2}+\left(  v_{y}^{0}\right)
^{2}\right]  }{2}=\frac{mR^{2}\omega^{2}}{2}. \label{1.14}%
\end{equation}
It is worth noting that in the limit of zero magnetic field, the $xy$-motion
tends to a uniform motion according to Eq. (\ref{1.6}), while diverges
according to Eqs. (\ref{1.13}), due to the $\omega^{-1}$ singularity. At a
first sight, this might be source of discrepancy but, to solve the Cauchy
problem, one has to specify the initial coordinates $\mathbf{x}_{0\perp}$ and
velocities $\mathbf{v}_{\perp}^{0}$. This means that the integration constants
$x_{c}$, $y_{c}$ and $\theta_{0}$ must be expressed in terms of the latter
constants. In fact such, solving Eqs. (\ref{1.12}) for $x_{c}$, $y_{c}$, and
$\theta_{0}$ and substituting into Eqs. (\ref{1.13}), the divergence is eliminated.

The quantum nonrelativistic motion of the system under consideration is
described by the corresponding Schr\"{o}dinger equation%
\begin{align}
&  i\hbar\partial_{t}\Psi\left(  \mathbf{r},t\right)  =\hat{H}\Psi\left(
\mathbf{r},t\right)  ,\ \ \hat{H}=\hat{H}_{xy}+\hat{H}_{z},\text{ \ }%
\partial_{t}=\frac{\partial}{\partial t},\nonumber\\
&  \hat{H}_{xy}=\frac{\mathbf{\hat{p}}_{\perp}^{2}}{2m}+\frac{m\omega^{2}}%
{8}\mathbf{r}_{\perp}^{2}-\frac{\omega}{2}\hat{L},\ \ \hat{L}=x\hat{p}%
_{y}-y\hat{p}_{x},\ \mathbf{\hat{p}}=-i\hbar\mathbf{\nabla},\nonumber\\
&  \hat{H}_{z}=\frac{\hat{p}_{z}^{2}}{2m}-m\xi z\sin^{2}\alpha+\xi\hat{p}%
_{z}t\cos^{2}\alpha+\frac{m\xi^{2}}{2}t^{2}\cos^{4}\alpha. \label{1.14a}%
\end{align}
Introducing dimensionless variables $\mathbf{q}$, $\boldsymbol{\hat{\pi}}$,
and $\tau$ as%
\begin{align}
&  \mathbf{q}=l^{-1}\mathbf{r}\,,\text{ \ }\tau=\frac{\hbar}{ml^{2}%
}t,\ \ \boldsymbol{\hat{\pi}}=\frac{l}{\hslash}\mathbf{\hat{p}}%
,\ \ \mathbf{\hat{\Pi}}=\frac{l}{\hbar}\mathbf{\hat{P}\rightarrow\hat{v}%
}=\frac{\hbar}{lm}\mathbf{\hat{\Pi},}\nonumber\\
&  \Omega=\frac{ml^{2}}{\hbar}\omega,\text{ \ }\Xi=\frac{m^{2}l^{3}}{\hbar
^{2}}\xi,\text{ \ }\left[  q_{k},\hat{\Pi}_{j}\right]  =\left[  q_{k},\hat
{\pi}_{j}\right]  =i\delta_{kj}, \label{1.15}%
\end{align}
we pass to an equivalent Schr\"{o}dinger equation%
\begin{align}
&  \hat{S}\Phi\left(  \mathbf{q},\tau\right)  =0,\text{ \ }\hat{S}%
=\partial_{\tau}+i\left(  \widehat{\mathcal{H}}_{\perp}+\widehat{\mathcal{H}%
}_{\parallel}\right)  ,\text{ \ }\Phi\left(  \mathbf{q},\tau\right)
=\sqrt{l^{3}}\Psi\left(  l\mathbf{q},\frac{ml^{2}}{\hbar}\tau\right)
,\nonumber\\
&  \widehat{\mathcal{H}}_{\perp}=\frac{\boldsymbol{\hat{\pi}}_{\bot}^{2}}%
{2}+\frac{\Omega^{2}\mathbf{q}_{\perp}^{2}}{8}-\frac{\Omega}{2}\mathcal{\hat
{L}},\text{ \ }\mathcal{\hat{L}}=q_{1}\hat{\pi}_{2}-q_{2}\hat{\pi}_{1},\text{
\ }\hat{H}_{xy}=\frac{\hbar^{2}}{ml^{2}}\widehat{\mathcal{H}}_{\perp
},\nonumber\\
&  \widehat{\mathcal{H}}_{\parallel}=\frac{\hat{\pi}_{3}^{2}}{2}-\Xi q_{3}%
\sin^{2}\alpha+\Xi\tau\hat{\pi}_{3}\cos^{2}\alpha+\frac{\Xi^{2}\tau^{2}}%
{2}\cos^{4}\alpha,\text{ \ }\hat{H}_{z}=\frac{\hbar^{2}}{ml^{2}}%
\widehat{\mathcal{H}}_{\parallel}. \label{1.16}%
\end{align}

\subsection{Special sets of integrals of motion}

Following well-known method \cite{Mal1979,Dod1987,Bag2015}, we are going to
find special sets $\hat{A}_{j}\left(  \tau\right)  $ of quantum integrals of
motion in the problem under consideration. They have to be linear combinations
of basic operators $\mathbf{q}$, $\boldsymbol{\hat{\pi}}$ and be, at the same
time, annihilation and creation operators$^{3}$\footnotetext[3]{The summation
convention over repeated scripts is assumed throughout, unless otherwise
explicitly stated.},%
\begin{align}
&  \hat{A}_{j}\left(  \tau\right)  =\frac{f_{jk}\left(  \tau\right)
q_{k}+ig_{jk}\left(  \tau\right)  \hat{\pi}_{k}}{\sqrt{2}}+\varphi_{j}\left(
\tau\right)  ,\label{2.1}\\
&  \left[  \hat{A}_{j}\left(  \tau\right)  ,\hat{A}_{k}^{\dagger}\left(
\tau\right)  \right]  =\delta_{jk},\text{ \ }\left[  \hat{A}_{j}\left(
\tau\right)  ,\hat{A}_{k}\left(  \tau\right)  \right]  =\left[  \hat{A}%
_{j}^{\dagger}\left(  \tau\right)  ,\hat{A}_{k}^{\dagger}\left(  \tau\right)
\right]  =0,\ \forall\tau. \label{2.2}%
\end{align}
Here, the coefficients $f_{jk}\left(  \tau\right)  $, $g_{jk}\left(
\tau\right)  $ and $\varphi_{j}\left(  \tau\right)  $ are some time-dependent functions.

One can easily see that both $\hat{A}_{j}\left(  \tau\right)  $ and $\hat
{A}_{j}^{\dagger}\left(  \tau\right)  $\ are integrals of motion if%
\begin{equation}
d_{\tau}\hat{A}_{j}\left(  \tau\right)  =\left[  \hat{S},\hat{A}_{j}\left(
\tau\right)  \right]  =0. \label{2.3}%
\end{equation}

Since the $xy$-motion is independent from the $z$-motion, the functions
$f_{3k}\left(  \tau\right)  $ and $g_{3k}\left(  \tau\right)  $ can be chosen
as%
\begin{equation}
f_{3k}\left(  \tau\right)  =f_{3}\left(  \tau\right)  \delta_{3k}%
,\ \ g_{3k}\left(  \tau\right)  =g_{3}\left(  \tau\right)  \delta_{3k},
\label{2.4}%
\end{equation}
which means that the operator $\hat{A}_{3}\left(  \tau\right)  $ depends on
the operators $q_{3}$ and $\hat{\pi}_{3}$ only. Under such a condition the
unknown functions $f_{jk}\left(  \tau\right)  $, $g_{jk}\left(  \tau\right)  $
and $\varphi_{j}\left(  \tau\right)  $ must obey the following equations%
\begin{align}
&  \dot{f}_{\beta1}\left(  \tau\right)  =\frac{\Omega}{2}f_{\beta2}\left(
\tau\right)  +\frac{i\Omega^{2}}{4}g_{\beta1}\left(  \tau\right)  ,\text{
\ }\dot{g}_{\beta1}\left(  \tau\right)  =if_{\beta1}\left(  \tau\right)
+\frac{\Omega}{2}g_{\beta2}\left(  \tau\right)  ,\text{ \ }\dot{\varphi
}_{\beta}\left(  \tau\right)  =0,\nonumber\\
&  \dot{f}_{\beta2}\left(  \tau\right)  =-\frac{\Omega}{2}f_{\beta1}\left(
\tau\right)  +\frac{i\Omega^{2}}{4}g_{\beta2}\left(  \tau\right)  ,\text{
\ }\dot{g}_{\beta2}\left(  \tau\right)  =if_{\beta2}\left(  \tau\right)
-\frac{\Omega}{2}g_{\beta1}\left(  \tau\right)  ,\nonumber\\
&  \dot{f}_{3}\left(  \tau\right)  =0,\text{ \ }\dot{g}\left(  \tau\right)
=if_{3}\left(  \tau\right)  ,\text{ \ }\dot{\varphi}_{3}\left(  \tau\right)
=\frac{\Xi\left[  g_{3}\left(  \tau\right)  \sin^{2}\alpha-i\tau f_{3}\left(
\tau\right)  \cos^{2}\alpha\right]  }{i\sqrt{2}}. \label{2.5}%
\end{align}

In addition, it follows from Eqs. (\ref{2.2}) and (\ref{2.3}) that functions
$f_{jk}\left(  \tau\right)  $ and $g_{jk}\left(  \tau\right)  $ are subjected
to the following conditions
\begin{align}
&  f_{jk}\left(  \tau\right)  g_{k^{\prime}k}^{\ast}\left(  \tau\right)
+f_{k^{\prime}k}^{\ast}\left(  \tau\right)  g_{jk}\left(  \tau\right)
=f_{jk}\left(  0\right)  g_{k^{\prime}k}^{\ast}\left(  0\right)
+f_{k^{\prime}k}^{\ast}\left(  0\right)  g_{jk}\left(  0\right)
=2\delta_{jk^{\prime}},\text{ }\nonumber\\
&  f_{jk}\left(  \tau\right)  g_{k^{\prime}k}\left(  \tau\right)
-f_{k^{\prime}k}\left(  \tau\right)  g_{jk}\left(  \tau\right)  =f_{jk}\left(
0\right)  g_{k^{\prime}k}\left(  0\right)  -f_{k^{\prime}k}\left(  0\right)
g_{jk}\left(  0\right)  =0. \label{2.6}%
\end{align}

The general solution of (\ref{2.5}) has the form%
\begin{align}
&  f_{\beta1}\left(  \tau\right)  =\frac{i\Omega}{2}c_{\beta1}-ib_{\beta
2}\frac{1+\cos\left(  \Omega\tau\right)  }{2}-\frac{ib_{\beta1}}{2}\sin\left(
\Omega\tau\right)  ,\text{ \ }\varphi_{\beta}\left(  \tau\right)
=\varphi_{\beta}\left(  0\right)  ,\nonumber\\
&  f_{\beta2}\left(  \tau\right)  =-\frac{i\Omega}{2}c_{\beta2}-ib_{\beta
1}\frac{1+\cos\left(  \Omega\tau\right)  }{2}+\frac{ib_{\beta2}}{2}\sin\left(
\Omega\tau\right)  ,\nonumber\\
&  g_{\beta1}\left(  \tau\right)  =c_{\beta2}+b_{\beta1}\frac{1-\cos\left(
\Omega\tau\right)  }{\Omega}+b_{\beta2}\frac{\sin\left(  \Omega\tau\right)
}{\Omega},\nonumber\\
&  g_{\beta2}\left(  \tau\right)  =c_{\beta1}-b_{\beta2}\frac{1-\cos\left(
\Omega\tau\right)  }{\Omega}+b_{\beta1}\frac{\sin\left(  \Omega\tau\right)
}{\Omega},\nonumber\\
&  f_{3}\left(  \tau\right)  =f_{0},\text{ \ }g_{3}\left(  \tau\right)
=g_{0}+if_{0}\tau,\text{ \ }\varphi_{3}\left(  \tau\right)  =-\left[
\frac{f_{0}\tau}{2}+ig_{3}\left(  \tau\right)  \sin^{2}\alpha\right]
\frac{\Xi\tau}{\sqrt{2}}, \label{2.7}%
\end{align}
where $c_{\beta\eta}$, $b_{\beta\eta}$, $f_{0}$ and $g_{0}$\ are some
constants. Without any loss of generality, we can set $\varphi_{j}\left(
0\right)  =0$.

It follows from (\ref{2.3}) that a common basis for $\hat{S}$ and $\hat{A}%
_{j}\left(  \tau\right)  $ can be found in the form%
\begin{align}
&  \hat{S}\left\vert \boldsymbol{\zeta},\tau\right\rangle =\lambda
_{\boldsymbol{\zeta}}\left(  \tau\right)  \left\vert \boldsymbol{\zeta}%
,\tau\right\rangle ,\label{2.8a}\\
&  \hat{A}_{j}\left(  \tau\right)  \left\vert \boldsymbol{\zeta}%
,\tau\right\rangle =\zeta_{j}\left\vert \boldsymbol{\zeta},\tau\right\rangle ,
\label{2.8b}%
\end{align}
wherein $\boldsymbol{\zeta}=\left(  \zeta_{1},\zeta_{2},\zeta_{3}\right)  $
are complex numbers and $\lambda_{\zeta}\left(  \tau\right)  $ is an arbitrary
time-dependent function.

\section{Generalized coherent states}

We define the GCS as solutions of the Schr\"{o}dinger equation that are
eigenstates of the operators $\hat{A}_{j}\left(  \tau\right)  $ with
eigenvalues $\zeta_{j}=\zeta_{\bot},\zeta_{3},$ $\zeta_{\bot}=\zeta_{1}%
,\zeta_{2}$. One can represent the GCS as%
\begin{equation}
\Phi_{\boldsymbol{\zeta}}\left(  \mathbf{q},\tau\right)  =\left\langle
\mathbf{q}|\boldsymbol{\zeta},\tau\right\rangle =\Phi_{\zeta_{\bot}}\left(
\mathbf{q}_{\bot},\tau\right)  \Phi_{\zeta_{3}}\left(  q_{3},\tau\right)  .
\label{c.1}%
\end{equation}
Then, to obey three conditions (\ref{2.8b}), the functions $\Phi_{\zeta_{\bot
}}\left(  \mathbf{q}_{\bot},\tau\right)  $ and $\Phi_{\zeta_{3}}\left(
q_{3},\tau\right)  $ have to satisfy the following equations%
\begin{align}
&  \hat{A}_{\beta}\Phi_{\zeta_{\bot}}\left(  \mathbf{q}_{\bot},\tau\right)
=\frac{f_{\beta1}q_{1}+f_{\beta2}q_{2}+g_{\beta1}\partial_{q_{1}}+g_{\beta
2}\partial_{q_{2}}}{\sqrt{2}}\Phi_{\zeta_{\bot}}\left(  \mathbf{q}_{\bot}%
,\tau\right)  =\zeta_{\beta}\Phi_{\zeta_{\bot}}\left(  \mathbf{q}_{\bot}%
,\tau\right)  ,\label{c.1a}\\
&  \hat{A}_{3}\Phi_{\zeta_{3}}\left(  q_{3},\tau\right)  =\left(  \frac
{f_{0}q_{3}+g_{3}\partial_{q_{3}}}{\sqrt{2}}+\varphi_{3}\right)  \Phi
_{\zeta_{3}}\left(  q_{3},\tau\right)  =\zeta_{3}\Phi_{\zeta_{3}}\left(
q_{3},\tau\right)  . \label{c.1b}%
\end{align}

The general solution of Eq. (\ref{c.1b}) reads%
\begin{equation}
\Phi_{\zeta_{3}}\left(  q_{3},\tau\right)  =\exp\left[  -\frac{f_{0}}{g_{3}%
}\frac{q_{3}^{2}}{2}+\sqrt{2}\frac{\zeta_{3}-\varphi_{3}}{g_{3}}q_{3}%
+i\phi_{3}\left(  \tau\right)  \right]  , \label{c.2}%
\end{equation}
where $\phi_{3}\left(  \tau\right)  $ is an arbitrary function of $\tau$.

The general solution of the set (\ref{c.1a}), see Appendix A, has the form%
\begin{equation}
\Phi_{\zeta_{\bot}}\left(  \mathbf{q}_{\bot},\tau\right)  =\exp\left[
-\frac{G_{1}q_{1}^{2}+G_{2}q_{2}^{2}}{2}+Fq_{1}q_{2}+\sqrt{2}\left(
Q_{1}q_{1}-Q_{2}q_{2}\right)  +i\phi_{\bot}\left(  \tau\right)  \right]  ,
\label{c.3}%
\end{equation}
where%
\begin{align}
G_{1}  &  =\frac{f_{11}g_{22}-f_{21}g_{12}}{g_{11}g_{22}-g_{12}g_{21}},\text{
\ }G_{2}=\frac{f_{22}g_{11}-f_{12}g_{21}}{g_{11}g_{22}-g_{12}g_{21}%
},\nonumber\\
Q_{1}  &  =\frac{g_{22}\zeta_{1}-g_{12}\zeta_{2}}{g_{11}g_{22}-g_{12}g_{21}%
},\text{ \ }Q_{2}=\frac{g_{21}\zeta_{1}-g_{11}\zeta_{2}}{g_{11}g_{22}%
-g_{12}g_{21}},\ F=\frac{f_{11}g_{21}-f_{21}g_{11}}{g_{11}g_{22}-g_{12}g_{21}%
}, \label{c.4}%
\end{align}
and $\phi_{\bot}\left(  \tau\right)  $ is an arbitrary time-dependent
function. Thus,%
\begin{align}
&  \Phi_{\boldsymbol{\zeta}}\left(  \mathbf{q},\tau\right)  =\Phi_{\zeta
_{\bot}}\left(  \mathbf{q}_{\bot},\tau\right)  \Phi_{\zeta_{3}}\left(
q_{3},\tau\right)  =\exp\left[  -\frac{G_{1}q_{1}^{2}+G_{2}q_{2}^{2}}%
{2}+Fq_{1}q_{2}\right. \nonumber\\
&  +\sqrt{2}\left(  Q_{1}q_{1}-Q_{2}q_{2}\right)  -\left.  \frac{f_{0}}{g_{3}%
}\frac{q_{3}^{2}}{2}+\sqrt{2}\frac{\zeta_{3}-\varphi_{3}}{g_{3}}q_{3}%
+i\phi\left(  \tau\right)  \right]  , \label{c.5}%
\end{align}
being $\phi\left(  \tau\right)  \equiv\phi_{3}\left(  \tau\right)  +\phi
_{\bot}\left(  \tau\right)  $ some function of $\tau$, such that
$\Phi_{\boldsymbol{\zeta}}\left(  \mathbf{q},\tau\right)  $ satisfies the
Schr\"{o}dinger equation (\ref{1.16}). This means that the function
$\phi\left(  \tau\right)  $ must be given by
\begin{align}
&  \phi\left(  \tau\right)  =\tilde{\phi}_{1}\left(  \tau\right)  +\frac{i}%
{2}\ln g_{3}-i\ln C,\nonumber\\
&  \tilde{\phi}_{1}\left(  \tau\right)  =\int\left[  Q_{1}^{2}+Q_{2}^{2}%
-\frac{G_{1}+G_{2}}{2}+\left(  \frac{\zeta_{3}-\varphi_{3}}{g_{3}}+\frac
{i\Xi\tau}{\sqrt{2}}\cos^{2}\alpha\right)  ^{2}\right]  d\tau, \label{c.6}%
\end{align}
where $C$ is a normalization constant, which can be found from the
normalization condition. The normalized function $\Phi_{\boldsymbol{\zeta}%
}\left(  \mathbf{q},\tau\right)  $ reads:%

\begin{align}
&  \Phi_{\boldsymbol{\zeta}}\left(  \mathbf{q},\tau\right)  =\left(
\frac{\operatorname{Re}G_{1}\operatorname{Re}G_{2}-\operatorname{Re}^{2}F}%
{\pi^{3}g_{3}^{2}}\right)  ^{1/4}\exp\left[  i\operatorname{Re}\tilde{\phi
}_{1}\left(  \tau\right)  -\left\vert g_{3}\right\vert ^{2}\operatorname{Re}%
^{2}\left(  \frac{\zeta_{3}-\varphi_{3}}{g_{3}}\right)  \right. \nonumber\\
&  -\frac{\operatorname{Re}G_{2}\operatorname{Re}^{2}Q_{1}-2\operatorname{Re}%
Q_{1}\operatorname{Re}Q_{2}\operatorname{Re}F+\operatorname{Re}G_{1}%
\operatorname{Re}^{2}Q_{2}}{\operatorname{Re}G_{1}\operatorname{Re}%
G_{2}-\operatorname{Re}^{2}F}\nonumber\\
&  \left.  -\frac{G_{1}q_{1}^{2}+G_{2}q_{2}^{2}}{2}-\frac{f_{0}}{g_{3}}%
\frac{q_{3}^{2}}{2}+Fq_{1}q_{2}+\sqrt{2}\left(  Q_{1}q_{1}-Q_{2}q_{2}\right)
+\sqrt{2}\frac{\zeta_{3}-\varphi_{3}}{g_{3}}q_{3}\right]  . \label{c.9}%
\end{align}
The family of states $\Phi_{\boldsymbol{\zeta}}\left(  \mathbf{q},\tau\right)
$ is parametrized by constants $c_{\beta\eta}$, $b_{\beta\eta}$, $f_{0}$ and
$g_{0}$.

It is possible to see from Eqs. (\ref{2.5}) that the following relations hold
\begin{equation}
g_{12}\left(  \tau\right)  =ig_{11}\left(  \tau\right)  \Longleftrightarrow
f_{12}\left(  \tau\right)  =if_{11}\left(  \tau\right)  \text{ \ }\&\text{
\ }g_{21}\left(  \tau\right)  =ig_{22}\left(  \tau\right)  \Longleftrightarrow
f_{21}\left(  \tau\right)  =if_{22}\left(  \tau\right)  . \label{c.10}%
\end{equation}
They imply, in turn, relations
\begin{equation}
c_{12}=-ic_{11},\text{ \ }c_{21}=-ic_{22},\text{ \ }b_{12}=-ib_{11},\text{
\ }b_{21}=-ib_{22}, \label{c.11}%
\end{equation}
and%
\begin{align}
&  f_{11}\left(  \tau\right)  =\frac{i\Omega}{2}c_{11}-b_{11}\frac
{1+e^{i\Omega\tau}}{2},\text{ \ }f_{22}\left(  \tau\right)  =-\frac{i\Omega
}{2}c_{22}-b_{22}\frac{1+e^{-i\Omega\tau}}{2},\nonumber\\
&  g_{11}\left(  \tau\right)  =-ic_{11}+b_{11}\frac{1-e^{i\Omega\tau}}{\Omega
},\text{ \ }g_{22}\left(  \tau\right)  =-ic_{22}-b_{22}\frac{1-e^{-i\Omega
\tau}}{\Omega}. \label{c.13}%
\end{align}
In addition, it follows from Eqs. (\ref{2.6}) that%
\begin{align}
&  2\operatorname{Re}\left(  f_{11}g_{11}^{\ast}\right)  =2\operatorname{Im}%
\left(  b_{11}c_{11}^{\ast}\right)  -\Omega\left\vert c_{11}\right\vert
^{2}=1,\text{ }\operatorname{Re}\left(  f_{0}g_{3}^{\ast}\right)
=\operatorname{Re}\left(  f_{0}g_{0}^{\ast}\right)  =1,\text{\ }\nonumber\\
&  2\operatorname{Re}\left(  f_{22}g_{22}^{\ast}\right)  =2\operatorname{Im}%
\left(  b_{22}c_{22}^{\ast}\right)  +\Omega\left\vert c_{22}\right\vert
^{2}=1,\text{ \ }\nonumber\\
&  f_{11}g_{22}-f_{22}g_{11}=ib_{22}c_{11}-\left(  ib_{11}+\Omega
c_{11}\right)  c_{22}=0. \label{c.12}%
\end{align}
On this stage, the GCS $\Phi_{\boldsymbol{\zeta}}\left(  \mathbf{q}%
,\tau\right)  $ can be rewritten as follows%
\begin{align}
\Phi_{\boldsymbol{\zeta}}\left(  \mathbf{q},\tau\right)   &  =\frac{1}%
{\pi^{3/4}\sqrt{2g_{11}^{2}g_{3}}}\exp\left[  i\operatorname{Re}\tilde{\phi
}_{2}\left(  \tau\right)  -2\left\vert g_{11}\right\vert ^{2}\left(
\operatorname{Re}^{2}Q_{1}+\operatorname{Re}^{2}Q_{2}\right)  -\left\vert
g_{3}\right\vert ^{2}\operatorname{Re}^{2}\left(  \frac{\zeta_{3}-\varphi_{3}%
}{g_{3}}\right)  \right. \nonumber\\
&  +\frac{i\Omega\tau}{2}-\left.  \frac{f_{11}}{g_{11}}\frac{q_{1}^{2}%
+q_{2}^{2}}{2}-\frac{f_{0}}{g_{3}}\frac{q_{3}^{2}}{2}+\sqrt{2}\frac{\zeta
_{3}-\varphi_{3}}{g_{3}}q_{3}+\sqrt{2}\left(  Q_{1}q_{1}-Q_{2}q_{2}\right)
\right]  , \label{c.14}%
\end{align}
where
\begin{align}
&  \tilde{\phi}_{2}\left(  \tau\right)  =\int\left[  Q_{1}^{2}+Q_{2}%
^{2}+\left(  \frac{\zeta_{3}-\varphi_{3}}{g_{3}}+\frac{i\Xi\tau}{\sqrt{2}}%
\cos^{2}\alpha\right)  ^{2}\right]  d\tau,\nonumber\\
&  Q_{1}=\frac{g_{22}\zeta_{1}-ig_{11}\zeta_{2}}{2g_{11}g_{22}},\text{
\ }Q_{2}=\frac{ig_{22}\zeta_{1}-g_{11}\zeta_{2}}{2g_{11}g_{22}}. \label{c.15}%
\end{align}

It follows from Eqs. (\ref{2.1}), (\ref{2.8b}) and (\ref{c.10}) that
\begin{align}
&  \zeta_{1}=\frac{f_{11}\left(  \tau\right)  \left[  q_{1}\left(
\tau\right)  +iq_{2}\left(  \tau\right)  \right]  +ig_{11}\left(  \tau\right)
\left[  \pi_{1}\left(  \tau\right)  +i\pi_{2}\left(  \tau\right)  \right]
}{\sqrt{2}},\nonumber\\
&  \zeta_{2}=\frac{if_{22}\left(  \tau\right)  \left[  q_{1}\left(
\tau\right)  -iq_{2}\left(  \tau\right)  \right]  -g_{22}\left(  \tau\right)
\left[  \pi_{1}\left(  \tau\right)  -i\pi_{2}\left(  \tau\right)  \right]
}{\sqrt{2}},\nonumber\\
&  \zeta_{3}=\frac{f_{0}q_{3}\left(  \tau\right)  +ig_{3}\left(  \tau\right)
\pi_{3}\left(  \tau\right)  }{\sqrt{2}}+\varphi_{3}\left(  \tau\right)  ,
\label{c.16a}%
\end{align}
wherein the mean values $q_{j}\left(  \tau\right)  =\left\langle
\boldsymbol{\zeta},\tau\left\vert q_{j}\right\vert \boldsymbol{\zeta}%
,\tau\right\rangle $ and$\ \pi_{j}\left(  \tau\right)  =\left\langle
\boldsymbol{\zeta},\tau\left\vert \hat{\pi}_{j}\right\vert \boldsymbol{\zeta
},\tau\right\rangle $ obey the classical equations of motion (\ref{1.4})
written in dimensionless variables,%
\begin{align}
\dot{\pi}_{1}\left(  \tau\right)   &  =\frac{\Omega}{2}\pi_{2}\left(
\tau\right)  -\frac{\Omega^{2}q_{1}\left(  \tau\right)  }{4},\text{ \ }%
\dot{\pi}_{2}\left(  \tau\right)  =-\frac{\Omega}{2}\pi_{1}\left(
\tau\right)  -\frac{\Omega^{2}q_{2}\left(  \tau\right)  }{4},\nonumber\\
\dot{q}_{1}\left(  \tau\right)   &  =\frac{\Omega}{2}q_{2}\left(  \tau\right)
+\pi_{1}\left(  \tau\right)  ,\text{ \ }\dot{q}_{2}\left(  \tau\right)
=-\frac{\Omega}{2}q_{1}\left(  \tau\right)  +\pi_{2}\left(  \tau\right)
,\nonumber\\
\dot{\pi}_{3}\left(  \tau\right)   &  =\Xi\sin^{2}\alpha,\text{ \ }\dot{q}%
_{3}\left(  \tau\right)  =\pi_{3}\left(  \tau\right)  +\Xi\tau\cos^{2}\alpha.
\label{c.17}%
\end{align}
In terms of classical trajectories $q_{j}\left(  \tau\right)  $ and $\pi
_{j}\left(  \tau\right)  $, the functions $\Phi_{\boldsymbol{\zeta}}\left(
\mathbf{q},\tau\right)  $ take the form%
\begin{align}
&  \Phi_{\boldsymbol{\zeta}}\left(  \mathbf{q},\tau\right)  =\frac{1}%
{\pi^{3/4}g_{11}\sqrt{2g_{3}}}\exp\left[  -\frac{f_{11}}{g_{11}}\frac{\left[
q_{1}-q_{1}\left(  \tau\right)  \right]  ^{2}+\left[  q_{2}-q_{2}\left(
\tau\right)  \right]  ^{2}}{2}-\frac{f_{0}}{g_{3}}\frac{\left[  q_{3}%
-q_{3}\left(  \tau\right)  \right]  ^{2}}{2}\right. \nonumber\\
&  -\left.  \frac{\pi_{1}\left(  \tau\right)  \left[  2q_{1}-q_{1}\left(
\tau\right)  \right]  +\pi_{2}\left(  \tau\right)  \left[  2q_{2}-q_{2}\left(
\tau\right)  \right]  +\pi_{3}\left(  \tau\right)  \left[  2q_{3}-q_{3}\left(
\tau\right)  \right]  }{2i}+\frac{i\Omega\tau}{2}+i\tilde{\phi}_{3}\right]  ,
\label{c.18}%
\end{align}
where%
\begin{equation}
\tilde{\phi}_{3}\left(  \tau\right)  =\tilde{\phi}_{3}^{\ast}\left(
\tau\right)  =\Xi\int\frac{q_{3}\left(  \tau\right)  \sin^{2}\alpha
-\tau\left(  \pi_{3}\left(  \tau\right)  +\Xi\tau\cos^{2}\alpha\right)
\cos^{2}\alpha}{2}d\tau. \label{c.19}%
\end{equation}

Let us consider a different representation for the GCS. Using the operators
$\hat{A}_{j}\left(  \tau\right)  $ and $\hat{A}_{j}^{\dagger}\left(
\tau\right)  $, we can construct a Fock space $\left\vert \mathbf{n}%
,\tau\right\rangle $,$\ \mathbf{n}=\left(  n_{1},n_{2},n_{3}\right)
$,$\ n_{j}=0,1,2,...,\ $at any time instant $\tau$,%
\begin{align}
&  \left\vert \mathbf{n},\tau\right\rangle =\frac{\left[  \hat{A}_{1}%
^{\dagger}\left(  \tau\right)  \right]  ^{n_{1}}\left[  \hat{A}_{2}^{\dagger
}\left(  \tau\right)  \right]  ^{n_{2}}\left[  \hat{A}_{3}^{\dagger}\left(
\tau\right)  \right]  ^{n_{3}}}{\sqrt{n_{1}!n_{2}!n_{3}!}}\left\vert
\mathbf{0},\tau\right\rangle ,\ \ \hat{A}_{j}\left(  \tau\right)  \left\vert
\mathbf{0},\tau\right\rangle =0,\nonumber\\
&  \left\langle \mathbf{m},\tau|\mathbf{n},\tau\right\rangle =\delta
_{m_{j}n_{j}}\,,\text{ \ }%
%TCIMACRO{\dsum \limits_{n_{1},n_{2},n_{3}=0}^{\infty}}%
%BeginExpansion
{\displaystyle\sum\limits_{n_{1},n_{2},n_{3}=0}^{\infty}}
%EndExpansion
\left\vert \mathbf{n},\tau\right\rangle \left\langle \mathbf{n},\tau
\right\vert =1. \label{c.21}%
\end{align}
Then,%
\begin{equation}
\left\vert \boldsymbol{\zeta},\tau\right\rangle =\exp\left(  -\frac{\left\vert
\boldsymbol{\zeta}\right\vert ^{2}}{2}\right)
%TCIMACRO{\dsum \limits_{n_{1},n_{2},n_{3}=0}^{\infty}}%
%BeginExpansion
{\displaystyle\sum\limits_{n_{1},n_{2},n_{3}=0}^{\infty}}
%EndExpansion
\frac{\zeta_{1}^{n_{1}}\zeta_{2}^{n_{2}}\zeta_{3}^{n_{3}}}{\sqrt{n_{1}%
!n_{2}!n_{3}!}}\left\vert \mathbf{n},\tau\right\rangle =\hat{D}\left(
\boldsymbol{\zeta},\tau\right)  \left\vert \mathbf{0},\tau\right\rangle ,
\label{c.22}%
\end{equation}
where the displacement operator $\hat{D}\left(  \boldsymbol{\zeta}%
,\tau\right)  $ reads%
\begin{equation}
\hat{D}\left(  \boldsymbol{\zeta},\tau\right)  \equiv e^{-\frac{\left\vert
\boldsymbol{\zeta}\right\vert ^{2}}{2}}e^{\zeta_{j}\hat{A}_{j}^{\dagger
}\left(  \tau\right)  }e^{-\zeta_{j}^{\ast}\hat{A}_{j}\left(  \tau\right)
}=e^{\zeta_{j}\hat{A}_{j}^{\dagger}\left(  \tau\right)  -\zeta_{j}^{\ast}%
\hat{A}_{j}\left(  \tau\right)  }. \label{c.23}%
\end{equation}
Using the completeness property of the states $\left\vert \mathbf{n}%
,\tau\right\rangle $, we can find their overlapping and prove a completeness
relation for the GCS,%
\begin{align}
&  \left\langle \boldsymbol{\zeta}^{\prime},\tau|\boldsymbol{\zeta}%
,\tau\right\rangle =\exp\left[  \left(  \boldsymbol{\zeta}^{\prime}\right)
^{\ast}\cdot\boldsymbol{\zeta}-\frac{\left\vert \boldsymbol{\zeta}^{\prime
}\right\vert ^{2}+\left\vert \boldsymbol{\zeta}\right\vert ^{2}}{2}\right]
,\text{ \ }\frac{1}{\pi^{3}}\int\int\left\vert \boldsymbol{\zeta}%
,\tau\right\rangle \left\langle \boldsymbol{\zeta},\tau\right\vert d^{2}%
\zeta=1,\nonumber\\
&  d^{2}\zeta=d^{2}\zeta_{1}d^{2}\zeta_{2}d^{2}\zeta_{3},\text{ \ }d^{2}%
\zeta_{j}=d\operatorname{Re}\zeta_{j}d\operatorname{Im}\zeta_{j},\text{
\ }\forall\tau. \label{c.24}%
\end{align}

The states $\left\vert \mathbf{n},\tau\right\rangle $ (\ref{c.21}) in
$\mathbf{q}$-representation $\Phi_{n}\left(  \mathbf{q},\tau\right)
=\left\langle \mathbf{q}|\mathbf{n},\tau\right\rangle $ have the form (see
appendix B)%
\begin{align}
&  \Phi_{n}\left(  \mathbf{q},\tau\right)  =\frac{1}{\sqrt{n_{1}!n_{2}!n_{3}%
!}}\left(  \frac{g_{3}^{\ast}}{2g_{3}}\right)  ^{\frac{n_{3}}{2}}H_{n_{3}%
}\left(  \frac{q_{3}+\sqrt{2}\operatorname{Re}\left(  \varphi_{3}g_{3}^{\ast
}\right)  }{\left\vert g_{3}\right\vert }\right) \nonumber\\
&  \times\left(  \frac{g_{11}^{\ast}}{g_{11}}\right)  ^{n_{1}}\left(
\frac{g_{22}^{\ast}}{ig_{11}^{\ast}}\right)  ^{n_{2}}H_{n_{1},n_{2}}\left(
\frac{q_{1}-iq_{2}}{\sqrt{2}g_{11}^{\ast}},\frac{q_{1}+iq_{2}}{\sqrt{2}g_{11}%
}\right)  \Phi_{0}\left(  \mathbf{q},\tau\right)  . \label{c.25}%
\end{align}

Using the relations%
\begin{align}
&  \exp\left(  \mu Z+\nu Z^{\ast}-\mu\nu\right)  =%
%TCIMACRO{\dsum \limits_{m_{1},m_{2}=0}^{\infty}}%
%BeginExpansion
{\displaystyle\sum\limits_{m_{1},m_{2}=0}^{\infty}}
%EndExpansion
H_{m_{1},m_{2}}\left(  Z,Z^{\ast}\right)  \frac{\mu^{m_{1}}\nu^{m_{2}}}%
{m_{1}!m_{2}!},\nonumber\\
&  \exp\left(  2\varkappa h-h^{2}\right)  =%
%TCIMACRO{\dsum \limits_{m=0}^{\infty}}%
%BeginExpansion
{\displaystyle\sum\limits_{m=0}^{\infty}}
%EndExpansion
\frac{H_{m}\left(  \varkappa\right)  }{m!}h^{m}, \label{c.26}%
\end{align}
we can calculate the sum in Eq. (\ref{c.22}) and rewrite the CS as follows%
\begin{align}
&  \Phi_{\boldsymbol{\zeta}}\left(  \mathbf{q},\tau\right)  =\Phi_{0}\left(
\mathbf{q},\tau\right)  \exp\left[  \sqrt{2}\frac{q_{3}+\sqrt{2}%
\operatorname{Re}\left(  \varphi_{3}g_{3}^{\ast}\right)  }{g_{3}}\zeta
_{3}-\frac{\zeta_{3}^{2}g_{3}^{\ast}}{2g_{3}}\right. \nonumber\\
&  +\left.  \zeta_{1}\frac{q_{1}-iq_{2}}{\sqrt{2}g_{11}}+\frac{\zeta_{2}%
g_{22}^{\ast}}{\left\vert g_{11}\right\vert ^{2}}\frac{q_{1}+iq_{2}}{i\sqrt
{2}}-\frac{g_{22}^{\ast}}{ig_{11}}\zeta_{1}\zeta_{2}-\frac{\left\vert
\boldsymbol{\zeta}\right\vert ^{2}}{2}\right]  , \label{c.27}%
\end{align}
where the vacuum state $\Phi_{0}\left(  \mathbf{q},\tau\right)  $ reads%
\begin{align}
&  \Phi_{0}\left(  \mathbf{q},\tau\right)  =\frac{1}{\pi^{3/4}g_{11}%
\sqrt{2g_{3}}}\exp\left[  i\operatorname{Re}\tilde{\phi}_{4}\left(
\tau\right)  -\left\vert g_{3}\right\vert ^{2}\operatorname{Re}^{2}\left(
\frac{\varphi_{3}}{g_{3}}\right)  +\frac{i\Omega\tau}{2}\right. \nonumber\\
&  \left.  -\frac{f_{11}}{g_{11}}\frac{q_{1}^{2}+q_{2}^{2}}{2}-\frac{f_{0}%
}{g_{3}}\frac{q_{3}^{2}}{2}-\sqrt{2}\frac{\varphi_{3}}{g_{3}}q_{3}\right]
,\ \ \tilde{\phi}_{4}\left(  \tau\right)  =\int\left(  \frac{i\Xi\tau}%
{\sqrt{2}}\cos^{2}\alpha-\frac{\varphi_{3}}{g_{3}}\right)  ^{2}d\tau.
\label{c.28}%
\end{align}
Using the relations%
\begin{equation}
\left\vert f_{11}\left(  \tau\right)  \right\vert =\left\vert f_{22}\left(
\tau\right)  \right\vert ,\text{ \ }\left\vert g_{11}\left(  \tau\right)
\right\vert =\left\vert g_{22}\left(  \tau\right)  \right\vert , \label{c.29}%
\end{equation}
one can easily verify that the states (\ref{c.27}) coincide with ones
(\ref{c.14}).

\subsection{Standard deviations and uncertainty relations}

We recall that the standard deviation $\sigma_{\chi}\left(  \tau\right)  $ of
a some physical quantity $\chi$ in the states $\left\vert \boldsymbol{\zeta
},\tau\right\rangle $ is calculated via the corresponding operator $\hat{\chi
}$ as follows%
\begin{align}
&  \sigma_{\chi}\left(  \tau\right)  \equiv\sqrt{\left\langle
\boldsymbol{\zeta},\tau|\left(  \hat{\chi}-\left\langle \boldsymbol{\zeta
},\tau|\hat{\chi}|\boldsymbol{\zeta},\tau\right\rangle \right)  ^{2}%
|\boldsymbol{\zeta},\tau\right\rangle }=\sqrt{\chi^{2}\left(  \tau\right)
-\left(  \chi\left(  \tau\right)  \right)  ^{2}},\nonumber\\
&  \left(  \chi\left(  \tau\right)  \right)  ^{2}\equiv\left(  \left\langle
\boldsymbol{\zeta},\tau|\hat{\chi}|\boldsymbol{\zeta},\tau\right\rangle
\right)  ^{2},\text{ \ }\chi^{2}\left(  \tau\right)  \equiv\left\langle
\boldsymbol{\zeta},\tau|\hat{\chi}^{2}|\boldsymbol{\zeta},\tau\right\rangle .
\label{D.1}%
\end{align}
Below, we calculate standard deviations of some physical quantities of the
problem under discussion. To this end, it is convenient to express the
operators $q_{j}$ and $\hat{\pi}_{j}$ in terms of the annihilation and
creation operators $\hat{A}_{j}\left(  \tau\right)  $ and $\hat{A}%
_{j}^{\dagger}\left(  \tau\right)  $. It follows from (\ref{2.1}), (\ref{2.2})
and (\ref{c.10}) that
\begin{align}
&  q_{1}=\frac{g_{11}^{\ast}\hat{A}_{1}+g_{11}\hat{A}_{1}^{\dagger}-i\left(
g_{22}^{\ast}\hat{A}_{2}-g_{22}\hat{A}_{2}^{\dagger}\right)  }{\sqrt{2}%
},\text{ \ }\hat{\pi}_{1}=\frac{f_{11}^{\ast}\hat{A}_{1}-f_{11}\hat{A}%
_{1}^{\dagger}-i\left(  f_{22}^{\ast}\hat{A}_{2}+f_{22}\hat{A}_{2}^{\dagger
}\right)  }{i\sqrt{2}},\nonumber\\
&  q_{2}=\frac{g_{22}^{\ast}\hat{A}_{2}+g_{22}\hat{A}_{2}^{\dagger}-i\left(
g_{11}^{\ast}\hat{A}_{1}-g_{11}\hat{A}_{1}^{\dagger}\right)  }{\sqrt{2}%
},\text{ \ }\hat{\pi}_{2}=\frac{f_{22}^{\ast}\hat{A}_{2}-f_{22}\hat{A}%
_{2}^{\dagger}-i\left(  f_{11}^{\ast}\hat{A}_{1}+f_{11}\hat{A}_{1}^{\dagger
}\right)  }{i\sqrt{2}},\nonumber\\
&  q_{3}=\frac{g_{3}^{\ast}\hat{A}_{3}+g_{3}\hat{A}_{3}^{\dagger
}-2\operatorname{Re}\left(  g_{3}^{\ast}\varphi_{3}\right)  }{\sqrt{2}},\text{
\ }\hat{\pi}_{3}=\frac{f_{0}^{\ast}\hat{A}_{3}-f_{0}\hat{A}_{3}^{\dagger
}-2i\operatorname{Im}\left(  f_{0}^{\ast}\varphi_{3}\right)  }{i\sqrt{2}}.
\label{D.2}%
\end{align}
Then, using Eq. (\ref{2.8b}) we can easily to find%
\begin{align}
&  q_{1}\left(  \tau\right)  =\sqrt{2}\left[  \operatorname{Im}\left(
g_{22}^{\ast}\zeta_{2}\right)  +\operatorname{Re}\left(  g_{11}^{\ast}%
\zeta_{1}\right)  \right]  =q_{1}^{0}-\frac{\Pi_{2}\left(  \tau\right)
-\Pi_{2}^{0}}{\Omega},\nonumber\\
&  q_{2}\left(  \tau\right)  =\sqrt{2}\left[  \operatorname{Im}\left(
g_{11}^{\ast}\zeta_{1}\right)  +\operatorname{Re}\left(  g_{22}^{\ast}%
\zeta_{2}\right)  \right]  =q_{2}^{0}+\frac{\Pi_{1}\left(  \tau\right)
-\Pi_{1}^{0}}{\Omega},\nonumber\\
&  \pi_{1}\left(  \tau\right)  =\sqrt{2}\left[  \operatorname{Im}\left(
f_{11}^{\ast}\zeta_{1}\right)  -\operatorname{Re}\left(  f_{22}^{\ast}%
\zeta_{2}\right)  \right]  =\frac{\Pi_{1}\left(  \tau\right)  +\Pi_{1}%
^{0}-\Omega q_{2}^{0}}{2},\nonumber\\
&  \pi_{2}\left(  \tau\right)  =\sqrt{2}\left[  \operatorname{Im}\left(
f_{22}^{\ast}\zeta_{2}\right)  -\operatorname{Re}\left(  f_{11}^{\ast}%
\zeta_{1}\right)  \right]  =\frac{\Pi_{2}\left(  \tau\right)  +\Pi_{2}%
^{0}+\Omega q_{1}^{0}}{2},\nonumber\\
&  q_{3}\left(  \tau\right)  =\sqrt{2}\operatorname{Re}\left[  g_{3}^{\ast
}\left(  \zeta_{3}-\varphi_{3}\right)  \right]  =q_{0}+\Pi_{3}^{0}\tau
+\frac{\Xi}{2}\tau^{2},\text{ \ }\nonumber\\
&  \pi_{3}\left(  \tau\right)  =\sqrt{2}\operatorname{Im}\left[  f_{0}^{\ast
}\left(  \zeta_{3}-\varphi_{3}\right)  \right]  =\Pi_{3}^{0}+\Xi\tau\sin
^{2}\alpha, \label{D.3}%
\end{align}
where
\begin{align}
&  \Pi_{1}\left(  \tau\right)  =\Pi_{1}^{0}\cos\left(  \Omega\tau\right)
+\Pi_{2}^{0}\sin\left(  \Omega\tau\right)  ,\text{ \ }\Pi_{2}\left(
\tau\right)  =\Pi_{2}^{0}\cos\left(  \Omega\tau\right)  -\Pi_{1}^{0}%
\sin\left(  \Omega\tau\right)  ,\nonumber\\
&  q_{1}^{0}=\sqrt{2}\left[  \operatorname{Re}\left(  c_{22}^{\ast}\zeta
_{2}\right)  -\operatorname{Im}\left(  c_{11}^{\ast}\zeta_{1}\right)  \right]
,\text{ \ }\Pi_{2}^{0}=\sqrt{2}\left[  \operatorname{Re}\left(  b_{11}^{\ast
}\zeta_{1}\right)  -\operatorname{Im}\left(  b_{22}^{\ast}\zeta_{2}\right)
\right]  ,\nonumber\\
&  q_{2}^{0}=\sqrt{2}\left[  \operatorname{Re}\left(  c_{11}^{\ast}\zeta
_{1}\right)  -\operatorname{Im}\left(  c_{22}^{\ast}\zeta_{2}\right)  \right]
,\text{ \ }\Pi_{1}^{0}=\sqrt{2}\left[  \operatorname{Re}\left(  b_{22}^{\ast
}\zeta_{2}\right)  -\operatorname{Im}\left(  b_{11}^{\ast}\zeta_{1}\right)
\right]  ,\nonumber\\
&  q_{0}=\sqrt{2}\operatorname{Re}\left(  g_{3}^{\ast}\zeta_{3}\right)
,~\ \Pi_{3}^{0}=\sqrt{2}\operatorname{Im}\left(  f_{0}^{\ast}\zeta_{3}\right)
. \label{D.4}%
\end{align}
Mean values of the operators $q_{j}^{2}$ and $\hat{\pi}_{j}^{2}$ in the states
$\left\vert \boldsymbol{\zeta},\tau\right\rangle $ are given by%
\begin{align}
q_{\beta}^{2}\left(  \tau\right)   &  =\left(  q_{\beta}\left(  \tau\right)
\right)  ^{2}+\frac{\left\vert g_{11}\right\vert ^{2}+\left\vert
g_{22}\right\vert ^{2}}{2},\text{ \ }\pi_{\beta}^{2}\left(  \tau\right)
=\left(  \pi_{\beta}\left(  \tau\right)  \right)  ^{2}+\frac{\left\vert
f_{11}\right\vert ^{2}+\left\vert f_{22}\right\vert ^{2}}{2},\nonumber\\
q_{3}^{2}\left(  \tau\right)   &  =\left(  q_{3}\left(  \tau\right)  \right)
^{2}+\frac{\left\vert g_{3}\right\vert ^{2}}{2},\text{ \ }\pi_{3}^{2}\left(
\tau\right)  =\left(  \pi_{3}\left(  \tau\right)  \right)  ^{2}+\frac
{\left\vert f_{0}\right\vert ^{2}}{2}. \label{D.5}%
\end{align}
Thus, we can find standard deviations of the position $\sigma_{q_{j}}\left(
\tau\right)  $ and momentum $\sigma_{\pi_{j}}\left(  \tau\right)  $,
\begin{align}
\sigma_{q_{\beta}}\left(  \tau\right)   &  \equiv\sigma_{q}\left(
\tau\right)  =\sqrt{\frac{\left\vert g_{11}\right\vert ^{2}+\left\vert
g_{22}\right\vert ^{2}}{2}},\text{ \ }\sigma_{\pi_{\beta}}\left(  \tau\right)
\equiv\sigma_{\pi}\left(  \tau\right)  =\sqrt{\frac{\left\vert f_{11}%
\right\vert ^{2}+\left\vert f_{22}\right\vert ^{2}}{2}},\nonumber\\
\sigma_{q_{3}}\left(  \tau\right)   &  =\frac{\left\vert g_{3}\right\vert
}{\sqrt{2}},\text{ \ }\sigma_{\pi_{3}}\left(  \tau\right)  =\sigma_{\pi_{3}%
}\left(  0\right)  =\sigma_{\pi_{3}}=\frac{\left\vert f_{0}\right\vert }%
{\sqrt{2}}. \label{D.6}%
\end{align}

By mean from (\ref{c.12}) and (\ref{D.6}), the constants $c_{\beta\beta}$,
$b_{\beta\beta}$, $f_{0}$ and $g_{0}$ can be related to the initial standard
deviations $\sigma_{q_{j}}\left(  0\right)  \equiv\sigma_{q_{j}}$ and
$\sigma_{\pi_{j}}\left(  0\right)  \equiv\sigma_{\pi_{j}}$ in the form
\begin{align}
&  \left\vert c_{11}\right\vert =\left\vert c_{22}\right\vert =\sigma
_{q},\text{ \ }\left\vert b_{11}\right\vert =\sigma_{\pi}\sqrt{1+\frac
{2\Omega+\Omega^{2}\sigma_{q}^{2}}{4\sigma_{\pi}^{2}}},\text{ \ }\left\vert
b_{22}\right\vert =\sigma_{\pi}\sqrt{1-\frac{2\Omega-\Omega^{2}\sigma_{q}^{2}%
}{4\sigma_{\pi}^{2}}},\nonumber\\
&  \operatorname{Im}\left(  c_{11}b_{11}^{\ast}+c_{22}^{\ast}b_{22}\right)
=-\Omega\sigma_{q}^{2},\text{ \ }\operatorname{Re}\left(  c_{11}b_{11}^{\ast
}+c_{22}^{\ast}b_{22}\right)  =\sqrt{4\sigma_{q}^{2}\sigma_{\pi}^{2}%
-1},\nonumber\\
&  \cos\left(  \arg g_{0}-\arg f_{0}\right)  =\frac{1}{2\sigma_{\pi_{3}}%
\sigma_{q_{3}}},\text{ \ }\operatorname{Im}\left(  f_{0}^{\ast}g_{0}\right)
=\sqrt{4\sigma_{q_{3}}^{2}\sigma_{\pi_{3}}^{2}-1}. \label{D.8}%
\end{align}
Taking the relations (\ref{c.13}) and (\ref{D.8}) into account, we can write
(\ref{D.6}) as follows%
\begin{align}
&  \sigma_{q}\left(  \tau\right)  =\sigma_{q}\sqrt{\cos\left(  \Omega
\tau\right)  +\frac{4\sigma_{\pi}^{2}+\Omega^{2}\sigma_{q}^{2}}{2\sigma
_{q}^{2}}\frac{1-\cos\left(  \Omega\tau\right)  }{\Omega^{2}}+\sqrt
{4\sigma_{q}^{2}\sigma_{\pi}^{2}-1}\frac{\sin\left(  \Omega\tau\right)
}{\Omega\sigma_{q}^{2}}},\nonumber\\
&  \sigma_{\pi}\left(  \tau\right)  =\sigma_{\pi}\sqrt{1+\left(  \frac
{\Omega^{2}\sigma_{q}^{2}}{4\sigma_{\pi}^{2}}-1\right)  \frac{1-\cos\left(
\Omega\tau\right)  }{2}-\frac{\Omega}{4\sigma_{\pi}^{2}}\sqrt{4\sigma_{q}%
^{2}\sigma_{\pi}^{2}-1}\sin\left(  \Omega\tau\right)  },\nonumber\\
&  \sigma_{q_{3}}\left(  \tau\right)  =\sigma_{q_{3}}\sqrt{1+\sqrt
{4\sigma_{\pi_{3}}^{2}-\frac{1}{\sigma_{q_{3}}^{2}}}\tau+\frac{\sigma_{\pi
_{3}}^{2}}{\sigma_{q_{3}}^{2}}\tau^{2}}. \label{D.7}%
\end{align}

One can see that $\sigma_{q}\left(  \tau\right)  $, $\sigma_{\pi}\left(
\tau\right)  $ and $\sigma_{q_{3}}\left(  \tau\right)  $ are real functions
if
\begin{equation}
\sigma_{q}\sigma_{\pi}\geq\frac{1}{2},\text{ \ }\sigma_{q_{3}}\sigma_{\pi_{3}%
}\geq\frac{1}{2}, \label{D.9}%
\end{equation}
which correspond the Heisenberg uncertainty relation in $\tau=0$.

In what follows, we consider the conditions%
\begin{equation}
\sigma_{\pi}=\frac{1}{2\sigma_{q}},\text{ \ }\sigma_{\pi_{3}}=\frac{1}%
{2\sigma_{q_{3}}}. \label{D.10}%
\end{equation}
Thus, the Heisenberg uncertainty relation is minimal at $\tau=0$. For any
$\tau$, we have that
\begin{align}
&  \sigma_{q}\left(  \tau\right)  \sigma_{\pi}\left(  \tau\right)  =\frac
{1}{2}\sqrt{1+\frac{1-\left(  2-\Omega^{2}\sigma_{q}^{4}\right)  \Omega
^{2}\sigma_{q}^{4}}{4\Omega^{2}\sigma_{q}^{4}}\sin^{2}\left(  \Omega
\tau\right)  }\geq\frac{1}{2},\nonumber\\
&  \sigma_{q_{3}}\left(  \tau\right)  \sigma_{\pi_{3}}=\frac{1}{2}%
\sqrt{1+\frac{\tau^{2}}{4\sigma_{q_{3}}^{4}}}\geq\frac{1}{2}. \label{D.11}%
\end{align}
The product $\sigma_{q}\left(  \tau\right)  \sigma_{\pi}\left(  \tau\right)  $
is minimal for any $\tau$\ if
\begin{equation}
\sigma_{q}=\frac{1}{\sqrt{\Omega}}. \label{D.11a}%
\end{equation}

One can see that the product $\sigma_{q}\left(  \tau\right)  \sigma_{\pi
}\left(  \tau\right)  $ is limited from above for any $\tau$,%
\begin{equation}
\frac{1}{2}\leq\sigma_{q}\left(  \tau\right)  \sigma_{\pi}\left(  \tau\right)
\leq\frac{1}{4}\sqrt{\frac{\left(  2+\Omega^{2}\sigma_{q}^{4}\right)
\Omega^{2}\sigma_{q}^{4}+1}{\Omega^{2}\sigma_{q}^{4}}}. \label{D.12}%
\end{equation}

To obtain the Robertson-Schr\"{o}dinger relation \cite{Sch1930,Rob1930} we
need calculate the covariance%
\begin{equation}
\sigma_{\chi\kappa}\left(  \tau\right)  =\frac{\left\langle \left(  \hat{\chi
}-\left\langle \hat{\chi}\right\rangle \right)  \left(  \hat{\kappa
}-\left\langle \hat{\kappa}\right\rangle \right)  +\left(  \hat{\kappa
}-\left\langle \hat{\kappa}\right\rangle \right)  \left(  \hat{\chi
}-\left\langle \hat{\chi}\right\rangle \right)  \right\rangle }{2}.
\label{D.15}%
\end{equation}
Then,%
\begin{align}
&  \sigma_{q_{\beta}\pi_{\beta}}\left(  \tau\right)  =\left\langle \hat{\pi
}_{\beta}q_{\beta}\right\rangle -\left\langle \hat{\pi}_{\beta}\right\rangle
\left\langle q_{\beta}\right\rangle +\frac{i}{2}=\frac{f_{11}^{\ast}%
g_{11}+f_{22}^{\ast}g_{22}-1}{2i},\nonumber\\
&  \sigma_{q_{3}\pi_{3}}\left(  \tau\right)  =\left\langle \hat{\pi}_{3}%
q_{3}\right\rangle -\left\langle \hat{\pi}_{3}\right\rangle \left\langle
q_{3}\right\rangle +\frac{i}{2}=\frac{f_{0}^{\ast}g_{3}-1}{2i}. \label{D.16}%
\end{align}
Thus, we have that%

\begin{equation}
\sigma_{q_{j}}^{2}\left(  \tau\right)  \sigma_{\pi_{j}}^{2}\left(
\tau\right)  -\sigma_{q_{j}\pi_{j}}^{2}\left(  \tau\right)  =\frac{1}{4}.
\label{D.17}%
\end{equation}
The Robertson-Schr\"{o}dinger uncertainty relations are minimized for any
$\tau$, this means that the GCS are squeezed states, see e.g., \cite{Lou1987}.

Now, let us study mean values and standard deviations of the velocities,
\begin{equation}
\hat{\Pi}_{1}=\hat{\pi}_{1}+\frac{\Omega}{2}q_{2},\text{ \ }\hat{\Pi}_{2}%
=\hat{\pi}_{2}-\frac{\Omega}{2}q_{1},\text{ \ }\hat{\Pi}_{3}=\hat{\pi}_{3}%
+\Xi\tau\cos^{2}\alpha. \label{D.18}%
\end{equation}
It follows from (\ref{D.3}) that
\begin{equation}
\Pi_{1}\left(  \tau\right)  =\pi_{1}\left(  \tau\right)  +\frac{\Omega}%
{2}q_{2}\left(  \tau\right)  ,\text{ \ }\Pi_{2}\left(  \tau\right)  =\pi
_{2}\left(  \tau\right)  -\frac{\Omega}{2}q_{1}\left(  \tau\right)  ,\text{
\ }\Pi_{3}\left(  \tau\right)  =\pi_{3}\left(  \tau\right)  +\Xi\tau\cos
^{2}\alpha. \label{D.19}%
\end{equation}
Using relations (\ref{D.3}), (\ref{D.5}) and (\ref{D.6}), we can calculate
mean values of the operators $\hat{\Pi}_{1}^{2}$, $\hat{\Pi}_{2}^{2}$ and
$\hat{\Pi}_{3}^{2}\ ,$%
\begin{align}
\Pi_{\beta}^{2}\left(  \tau\right)   &  =\left(  \Pi_{\beta}\left(
\tau\right)  \right)  ^{2}+\sigma_{\pi}^{2}\left(  \tau\right)  +\frac
{\Omega^{2}\sigma_{q}^{2}\left(  \tau\right)  }{4}=\left(  \Pi_{\beta}\left(
\tau\right)  \right)  ^{2}+\frac{1+\Omega^{2}\sigma_{q}^{4}}{4\sigma_{q}^{2}%
},\text{ \ }\nonumber\\
\Pi_{3}^{2}\left(  \tau\right)   &  =\left(  \Pi_{3}\left(  \tau\right)
\right)  ^{2}+\frac{\left\vert f_{0}\right\vert ^{2}}{2}. \label{D.20}%
\end{align}
Thus, we find the Standard deviations of the velocities in the form%
\begin{equation}
\sigma_{\Pi_{\beta}}\left(  \tau\right)  =\frac{\sqrt{1+\Omega^{2}\sigma
_{q}^{4}}}{2\sigma_{q}}\equiv\sigma_{\Pi},\text{ \ }\sigma_{\Pi_{3}}\left(
\tau\right)  =\frac{\left\vert f_{0}\right\vert }{\sqrt{2}}\equiv\sigma
_{\pi_{3}}. \label{D.21}%
\end{equation}

Taking into account relation (\ref{D.10}), we see that in (\ref{c.18}) exist
families of GCS which differ one from another by values of the parameters
$\sigma=\sigma_{q}$,$\sigma_{q_{3}}$,%
\begin{align}
&  \Phi_{\sigma}\left(  \mathbf{q},\tau\right)  =\frac{\exp\left[
-\frac{f_{11}^{\sigma_{q}}\left(  \tau\right)  }{g_{11}^{\sigma_{q}}\left(
\tau\right)  }\frac{\left[  q_{1}-q_{1}\left(  \tau\right)  \right]
^{2}+\left[  q_{2}-q_{2}\left(  \tau\right)  \right]  ^{2}}{2}-\frac{\left[
q_{3}-q_{3}\left(  \tau\right)  \right]  ^{2}}{2\left(  2\sigma_{q_{3}}%
^{2}+i\tau\right)  }+i\tilde{\phi}_{3}+\frac{i\Omega\tau}{2}\right.  }%
{g_{11}^{\sigma_{q}}\left(  \tau\right)  \sqrt{\sqrt{2^{3}\pi^{3}}\left(
\sigma_{q_{3}}+\frac{i\tau}{2\sigma_{q_{3}}}\right)  }}\nonumber\\
&  \left.  -\frac{\pi_{1}\left(  \tau\right)  \left[  2q_{1}-q_{1}\left(
\tau\right)  \right]  +\pi_{2}\left(  \tau\right)  \left[  2q_{2}-q_{2}\left(
\tau\right)  \right]  +\pi_{3}\left(  \tau\right)  \left[  2q_{3}-q_{3}\left(
\tau\right)  \right]  }{2i}\right]  , \label{D.22}%
\end{align}
where%
\begin{equation}
f_{11}^{\sigma_{q}}\left(  \tau\right)  =\frac{i\Omega\sigma_{q}}{2}%
+\frac{\left(  1+\Omega\sigma_{q}^{2}\right)  \left(  1+e^{i\Omega\tau
}\right)  }{4i\sigma_{q}},\text{ \ }g_{11}^{\sigma_{q}}\left(  \tau\right)
=-i\sigma_{q}-\frac{\left(  1+\Omega\sigma_{q}^{2}\right)  \left(
1-e^{i\Omega\tau}\right)  }{2i\Omega\sigma_{q}}. \label{D.23}%
\end{equation}

Setting electric $\mathbf{E}$ and magnetic $\mathbf{B}$ fields to zero in Eq.
(\ref{D.22}), we obtain the GCS for the free $3$-dimensional particle,%
\begin{align}
&  \Phi_{\boldsymbol{\zeta}}\left(  \mathbf{q},\tau\right)  =\frac
{\exp\left\{  -\frac{\left[  q_{1}-q_{1}\left(  \tau\right)  \right]  ^{2}%
}{2\left(  2\sigma_{q}^{2}+i\tau\right)  }-\frac{\left[  q_{2}-q_{2}\left(
\tau\right)  \right]  ^{2}}{2\left(  2\sigma_{q}^{2}+i\tau\right)  }%
-\frac{\left[  q_{3}-q_{3}\left(  \tau\right)  \right]  ^{2}}{2\left(
2\sigma_{q_{3}}^{2}+i\tau\right)  }\right.  }{\sqrt{\sqrt{2^{3}\pi^{3}}\left(
\sigma_{q}+\frac{i\tau}{2\sigma_{q}}\right)  ^{2}\left(  \sigma_{q_{3}}%
+\frac{i\tau}{2\sigma_{q_{3}}}\right)  }}\nonumber\\
&  \left.  -\frac{\pi_{1}\left[  2q_{1}-q_{1}\left(  \tau\right)  \right]
+\pi_{2}\left[  2q_{2}-q_{2}\left(  \tau\right)  \right]  +\pi_{3}\left[
2q_{3}-q_{3}\left(  \tau\right)  \right]  }{2i}\right\}  , \label{F.2}%
\end{align}
where $q_{j}\left(  \tau\right)  =q_{j}^{0}+\Pi_{j}^{0}\tau$ and $\pi_{j}%
=\Pi_{j}^{0}$.

The function (\ref{F.2}) is a product of three one-dimensional GCS that depend
on $q_{1}$, $q_{2}\ $and $q_{3}$ respectively. These one-dimensional GCS
coincide with ones obtained and studied in our previous publications
\cite{Bag2014}.

\subsection{CS of a charged particle in constant magnetic field}

Quantum nonrelativistic motion of a charged particle in a constant magnetic
field was studied in a number of articles, see e.g.,
\cite{Dar1927,Lan1930,Joh1949,Sey1965,Ber1982}. CS of such a system were
obtained first by Malkin and Man'ko \cite{Mal1968}, see too
\cite{Mal1969,Dod1971,Var1984,Rem2005,Set2009}.

The states (\ref{c.27}) for zero electric field are GCS for such a system. To
compare the latter states with ones by Malkin and Man'ko one has to projet
them on the $xy$-plane and to use the relation (\ref{D.11a}). In Ref.
\cite{Mal1968}, the CS of $\hat{H}_{xy}$ (\ref{1.14a}) are given by $\left(
\hbar=1\right)  $%
\begin{equation}
\Phi_{\alpha\beta}\left(  \varepsilon\right)  =\sqrt{\frac{m\omega}{2\pi}}%
\exp\left(  -\left\vert \varepsilon\right\vert ^{2}+\sqrt{2}\beta
\varepsilon+i\sqrt{2}\alpha\varepsilon^{\ast}-i\alpha\beta-\frac{\left\vert
\alpha\right\vert ^{2}+\left\vert \beta\right\vert ^{2}}{2}\right)  ,
\label{A1}%
\end{equation}
where%
\begin{align}
&  \varepsilon=\sqrt{m\omega}\frac{x+iy}{2},\text{ \ }\hat{a}=-\frac{i}%
{\sqrt{2}}\left(  \varepsilon+\frac{\partial}{\partial\varepsilon^{\ast}%
}\right)  ,\text{ \ }\hat{b}=\frac{1}{\sqrt{2}}\left(  \varepsilon^{\ast
}+\frac{\partial}{\partial\varepsilon}\right) \nonumber\\
&  \hat{a}\Phi_{\alpha\beta}=\alpha\Phi_{\alpha\beta},\text{ \ }\hat{b}%
\Phi_{\alpha\beta}=\beta\Phi_{\alpha\beta}. \label{A2}%
\end{align}
Being rewritten in term of the operators $\hat{a}$ and $\hat{b}$ the
Hamiltonian $\hat{H}_{xy}$ takes the form%
\begin{equation}
\hat{H}_{xy}=\omega\left(  \hat{a}^{\dagger}\hat{a}+\frac{1}{2}\right)  .
\label{A3}%
\end{equation}
Then, the time evolution of $\Phi_{\alpha\beta}\left(  \varepsilon\right)  $
is given by means of $\alpha\left(  t\right)  =\alpha e^{-i\omega t}$ as
follows%
\begin{align}
\Phi_{\alpha\beta}\left(  \varepsilon,t\right)   &  =\exp\left(  -i\hat
{H}_{xy}t\right)  \Phi_{\alpha\beta}\left(  \varepsilon\right) \nonumber\\
&  =\sqrt{\frac{m\omega}{2\pi}}\exp\left[  -\left\vert \varepsilon\right\vert
^{2}+\sqrt{2}\left(  i\alpha e^{-i\omega t}\varepsilon^{\ast}+\beta
\varepsilon\right)  -i\alpha e^{-i\omega t}\beta-\frac{\left\vert
\alpha\right\vert ^{2}+\left\vert \beta\right\vert ^{2}}{2}-\frac{i\omega
t}{2}\right]  . \label{A4}%
\end{align}

Now, let us consider the states (\ref{c.27}) in dimensional variables
(\ref{1.15}), with $\hbar=1$, on the plane $xy$ as follows%
\begin{align}
\Psi_{\sigma_{\bot}}\left(  \mu,t\right)   &  =\frac{1}{\sqrt{\pi}%
g_{11}^{\sigma_{\bot}}}\exp\left[  -\frac{f_{11}^{\sigma_{\bot}}}%
{g_{11}^{\sigma_{\bot}}}\frac{\left\vert \mu\right\vert ^{2}}{2}+\frac
{1}{\sqrt{2}}\left(  \frac{\zeta_{1}\mu^{\ast}}{g_{11}^{\sigma_{q}}}%
+\frac{\zeta_{2}\mu}{ig_{22}}\right)  \right. \nonumber\\
&  \left.  -\frac{g_{22}^{\ast\sigma_{\bot}}}{ig_{11}^{\sigma_{\bot}}}%
\zeta_{1}\zeta_{2}-\frac{\left\vert \zeta_{1}\right\vert ^{2}+\left\vert
\zeta_{2}\right\vert ^{2}}{2}+\frac{i\omega t}{2}\right]  , \label{A5}%
\end{align}
where
\begin{align}
&  f_{11}^{\sigma_{\bot}}\left(  t\right)  =\frac{\left(  1+m\omega
\sigma_{\bot}^{2}\right)  \left(  1+e^{i\omega t}\right)  -2m\omega
\sigma_{\bot}^{2}}{4i\sigma_{\bot}},\text{ \ }f_{22}^{\sigma_{\bot}}\left(
t\right)  =\frac{2m\omega\sigma_{\bot}^{2}+\left(  1-m\omega\sigma_{\bot}%
^{2}\right)  \left(  1+e^{-i\omega t}\right)  }{4i\sigma_{\bot}},\nonumber\\
&  g_{11}^{\sigma_{\bot}}\left(  t\right)  =\frac{2m\omega\sigma_{\bot}%
^{2}-\left(  1+m\omega\sigma_{\bot}^{2}\right)  \left(  1-e^{i\omega
t}\right)  }{2im\omega\sigma_{\bot}},\text{ \ }g_{22}^{\sigma_{\bot}}\left(
t\right)  =\frac{2m\omega\sigma_{\bot}^{2}+\left(  1-m\omega\sigma_{\bot}%
^{2}\right)  \left(  1-e^{-i\omega t}\right)  }{2im\omega\sigma_{\bot}%
},\nonumber\\
&  \mu=x+iy,\text{ \ }\sigma_{q}=l^{-1}\sigma_{\bot},\text{ \ }\Phi
_{\boldsymbol{\zeta}_{\bot}}\left(  \mathbf{q}_{\bot},\tau\right)
=l\Psi_{\sigma_{\bot}}\left(  \mu,t\right)  . \label{A6}%
\end{align}
Using the condition (\ref{D.11a}) as follows%
\begin{equation}
\sigma_{\bot}=\frac{1}{\sqrt{m\omega}}. \label{A7}%
\end{equation}
We find that%
\begin{equation}
f_{11}^{\sigma_{\bot}}\left(  t\right)  =\frac{le^{i\omega t}\sqrt{m\omega}%
}{2i},\text{ \ }g_{11}^{\sigma_{\bot}}\left(  t\right)  =\frac{e^{i\omega t}%
}{il\sqrt{m\omega}},\text{ \ }f_{22}^{\sigma_{\bot}}=\frac{l\sqrt{m\omega}%
}{2i},\text{ \ }g_{22}^{\sigma_{\bot}}=\frac{1}{il\sqrt{m\omega}}. \label{A8}%
\end{equation}
Thus, the states (\ref{A5}), $\Psi_{\sigma_{\bot}}\left(  \mu,t\right)
\longrightarrow\Psi\left(  \mu,t\right)  $, take the form%
\begin{align}
\Psi\left(  \mu,t\right)   &  =\sqrt{\frac{m\omega}{\pi}}\exp\left[
-\frac{m\omega\left\vert \mu\right\vert ^{2}}{4}+\sqrt{\frac{m\omega}{2}%
}\left(  i\zeta_{1}\mu^{\ast}e^{i\omega t}+\zeta_{2}\mu\right)  \right.
\nonumber\\
&  \left.  -ie^{-i\omega t}\zeta_{1}\zeta_{2}-\frac{\left\vert \zeta
_{1}\right\vert ^{2}+\left\vert \zeta_{2}\right\vert ^{2}}{2}-\frac{i\omega
t}{2}\right]  . \label{A9}%
\end{align}
Using the relations%
\begin{equation}
\mu=\frac{2}{\sqrt{m\omega}}\varepsilon,\text{ \ }\zeta_{1}=\alpha,\text{
\ }\zeta_{2}=\beta, \label{A10}%
\end{equation}
we can show that%
\begin{equation}
\Psi\left(  \mu,t\right)  =\Phi_{\alpha\beta}\left(  \varepsilon,t\right)  .
\label{A11}%
\end{equation}
Thus, we find that the GCS (\ref{c.27}) coincides with the CS of Malkin and
Man'ko given by the condition (\ref{A7}).

\section{GCS as semiclassical states}

In general case, the GCS cannot be considered as semiclassical states (SS)
because the standard deviations can grow over the time or by means of any
other parameter of the system. The study of the SS have attracted attention of
theoretical physicists and chemical, see e.g.,
\cite{Alf1965,Hel1975,Hel1981,Lit1986,Gro1998,Mil2001,Bar2001,Bag2011,Ohs2013,Ozo2016}%
. We have that the GCS can considered as SS if the conditions below are satisfied:

\textbf{I} - Position and velocity mean value must propagate along the
classical trajectories.

\textbf{II} - Position and velocity standard deviation must have a short
interval of variation.

In our previous publication \cite{Bag2014}, we find that SS of a free particle
is given by condition%
\begin{equation}
v\gg\frac{\hbar}{2m\sigma_{x}}, \label{S.1}%
\end{equation}
wherein $v$ is the velocity and $\sigma_{x}$ is the position standard
deviation. This means that the velocity mean value must be much larger than
its corresponding standard deviation. Hence, the variation of the position
mean value is much larger than the variation of its corresponding standard
deviation. In a certain sense, this implies that the mean values propagate
with a constant standard deviation, which is consistent with the conditions
\textbf{I} and \textbf{II} listed above. Of course that in a long instant time
we will not have SS anymore.

First, let us study the conditions of SS on the $z$-motion. To this, we return
to the initial dimensional variables and rewrite the Eqs. (\ref{D.3})
(\ref{D.7}), (\ref{D.19}) and (\ref{D.21}) in these variables, as follows%
\begin{align*}
&  z\left(  t\right)  =z_{0}+v_{z}^{0}t+\frac{\xi}{2}t^{2},\text{ \ }%
v_{z}\left(  \tau\right)  =v_{z}^{0}+\xi t,\text{ \ }\\
&  \sigma_{z}\left(  t\right)  =\sigma_{z}\sqrt{1+\frac{\hbar^{2}}%
{4m^{2}\sigma_{z}^{4}}t^{2}},\text{ \ }\sigma_{v_{z}}=\frac{\hbar}%
{2m\sigma_{z}}.
\end{align*}
Consider the inequality between the velocity $v_{z}\left(  \tau\right)  $ and
its corresponding standard deviation $\sigma_{v_{z}}$ in the form
\begin{equation}
v_{z}^{0}+\xi t\gg\frac{\hbar}{2m\sigma_{z}}. \label{S.2}%
\end{equation}
One can see that at any time instant $t$ the inequality holds true provided
that the condition below is satisfied%
\begin{equation}
v_{z}^{0}\gg\frac{\hbar}{2m\sigma_{z}}. \label{S.3}%
\end{equation}
The condition (\ref{S.3}) implies that the variation of $z\left(  t\right)  $
is much larger than the variation of $\sigma_{z}\left(  t\right)  $. Thus, we
can write
\begin{equation}
\frac{\Delta z\left(  t\right)  }{\Delta\sigma_{z}\left(  t\right)  }%
=\frac{z\left(  t\right)  -z\left(  0\right)  }{\sigma_{z}\left(  t\right)
-\sigma_{z}\left(  0\right)  }\gg1\Rightarrow\frac{\frac{v_{z}^{0}}{\sigma
_{z}}t+\frac{\xi}{2\sigma_{z}}t^{2}}{\sqrt{1+\frac{\hbar^{2}}{4m^{2}\sigma
_{z}^{4}}t^{2}}}\gg1. \label{S.4}%
\end{equation}

The electric field allows to have SS when the condition (\ref{S.3}) is no
longer true. However, this is possible after an instant of time, given by%
\begin{equation}
t\gg\left(  \frac{\hbar}{2m\sigma_{z}}-v_{z}^{0}\right)  \frac{1}{\xi}.
\label{S.5}%
\end{equation}
We can see that much larger are the parameters $\xi$, $v_{z}^{0}$ is smaller
the time required for the conditions of SS to be satisfied. Then, it follows
from (\ref{S.4}) that%
\begin{equation}
\frac{\xi}{2\sigma_{z}}\gg\frac{\hbar^{2}}{4m^{2}\sigma_{z}^{4}}\Rightarrow
E\gg\frac{\hbar^{2}}{2me\sigma_{z}^{3}}. \label{S.6}%
\end{equation}

On the $xy$-plane, the standard deviation of the position $\sigma_{\bot
}\left(  t\right)  $ and velocity $\sigma_{v_{\bot}}$ are given by%
\begin{align}
&  \sigma_{x}\left(  t\right)  =\sigma_{y}\left(  t\right)  \equiv\sigma
_{\bot}\left(  t\right)  =\sigma_{\bot}\sqrt{\cos\left(  \omega t\right)
+\left(  1+\frac{\hbar^{2}}{m^{2}\omega^{2}\sigma_{\bot}^{4}}\right)
\frac{1-\cos\left(  \omega t\right)  }{2}},\nonumber\\
&  \sigma_{v_{x}}=\sigma_{v_{y}}\equiv\sigma_{v_{\bot}}=\frac{\hslash
}{2m\sigma_{\bot}}\sqrt{1+\frac{m^{2}\sigma_{\bot}^{4}\omega^{2}}{\hbar^{2}}}.
\label{S.7}%
\end{align}
In this case, the motion is limited and therefore a separate analysis must be
carried out beyond those discussed for $z$-motion. One can see that the
standard deviations depends on the magnetic field $B$. Thus, we must analysis
the following cases:

\textbf{Case I}%
\begin{equation}
\frac{m^{2}\sigma_{\bot}^{4}\omega^{2}}{\hbar^{2}}\gg1\Rightarrow B\gg
\frac{\hbar c}{e\sigma_{\bot}^{2}}, \label{S.8}%
\end{equation}
the velocity standard deviation becomes much large and the position standard
deviation becomes the smallest possible.

\textbf{Case II}
\begin{equation}
\frac{m^{2}\sigma_{\bot}^{4}\omega^{2}}{\hbar^{2}}=1\Rightarrow B=\frac{\hbar
c}{e\sigma_{\bot}^{2}}, \label{S.9}%
\end{equation}
the position standard deviation becomes a constant
\begin{equation}
\sigma_{\bot}\left(  t\right)  =\sigma_{\bot}\left(  0\right)  =\sigma_{\bot},
\label{S.10}%
\end{equation}
and the velocity standard deviation is given by%
\begin{equation}
\sigma_{v_{\bot}}=\frac{\hslash}{\sqrt{2}m\sigma_{\bot}}. \label{S.11}%
\end{equation}

\textbf{Case III}%
\begin{equation}
\frac{m^{2}\sigma_{\bot}^{4}\omega^{2}}{\hbar^{2}}=0\Rightarrow B=0,
\label{S.12}%
\end{equation}
we have that the velocity standard deviation assumes its lowest value%
\begin{equation}
\sigma_{v_{\bot}}=\frac{\hslash}{2m\sigma_{\bot}}. \label{S.13}%
\end{equation}
But, the position standard deviation begins to grow over time%
\begin{equation}
\sigma_{\bot}\left(  t\right)  =\sigma_{\bot}\sqrt{1+\frac{\hbar^{2}}%
{4m^{2}\sigma_{\bot}^{4}}t^{2}}. \label{S.14}%
\end{equation}

\textbf{Case IV}%
\begin{equation}
0<\frac{m^{2}\sigma_{\bot}^{4}\omega^{2}}{\hbar^{2}}<1\Rightarrow
0<B<\frac{\hbar c}{e\sigma_{\bot}^{2}}, \label{S.15}%
\end{equation}
The simultaneous measurements in the position and velocity have the standard
deviation with the lowest value possible. In addition, we must impose the
condition below, such that in the limit of null field the conditions of SS
holds true
\begin{equation}
v_{x}^{0}\gg\frac{\hbar}{2m\sigma_{\bot}}\sqrt{1+\frac{m^{2}\sigma_{\bot}^{4}%
}{\hbar^{2}}\omega^{2}},\text{ \ }v_{y}^{0}\gg\frac{\hbar}{2m\sigma_{\bot}%
}\sqrt{1+\frac{m^{2}\sigma_{\bot}^{4}}{\hbar^{2}}\omega^{2}}. \label{S.16}%
\end{equation}

We can see that the condition of SS is not satisfied only in \textbf{Case I}
because the standard deviation of the velocity is a function of the magnetic
field. The \textbf{Case IV} presents the best description of the SS because
the position and velocity mean values presented simultaneously the lowest
value in their measurements. In addition, it admit the free particle limit.

Taking into account the limit $E=B=0$, we find the conditions of SS of a free
particle in 3-dimension from (\ref{S.2}) and (\ref{S.16}) as follows
\begin{equation}
v_{x}^{0}\gg\frac{\hbar}{2m\sigma_{x}},\text{ \ }v_{y}^{0}\gg\frac{\hbar
}{2m\sigma_{y}},\text{ \ }v_{z}^{0}\gg\frac{\hbar}{2m\sigma_{z}}. \label{S.17}%
\end{equation}

\section{Concluding remarks}

In this work, we obtain the classical equations of motion which are reduced to
the free particle case in the limit of null field by mean of the initial
Cauchy data. We find the corresponding GCS that are solutions of the
Schr\"{o}dinger equation, parametrized by the standard deviations $\sigma_{x}%
$, $\sigma_{y}$ and $\sigma_{z}$ at the initial time instant. These states are
squeezed states which present potential application in optical. In addition,
we can obtain the GCS of a free particle taking into account the limit of null
field, such result it is important to the scattering process. We obtain
conditions under which the GCS can be considered as SS and these conditions
hold on the limit of null field.

\begin{acknowledgments}
This work is supported by the Tomsk Polytechnic University Competitiveness
Enhancement Program under Grant VIU-FTI-72/2017.
\end{acknowledgments}

\appendix

\section{Solution from the set (\ref{c.1a})}

To solve equations from the set (\ref{c.1a}), we are going to use the fact
(see Ref. \cite{Pol2002}) that%
\begin{equation}
w\left(  x,y\right)  =\Upsilon\left(  z\right)  \exp\left[  \frac{1}{a}\int
f\left(  x\right)  dx+\frac{1}{b}\int g\left(  y\right)  dy\right]
,\ z=bx-ay, \label{A.1}%
\end{equation}
with an arbitrary function $\Upsilon\left(  z\right)  $ is the general
solution of the equation
\begin{equation}
\left(  a\partial_{x}+b\partial_{y}\right)  w\left(  x,y\right)  =\left[
f\left(  x\right)  +g\left(  y\right)  \right]  w\left(  x,y\right)  .
\label{A.2}%
\end{equation}
Consider the Eq. (\ref{c.1b}) with $\beta=2$,
\begin{equation}
\left(  g_{21}\partial_{q_{1}}+g_{22}\partial_{q_{2}}\right)  \Phi
_{\zeta_{\bot}}\left(  \mathbf{q}_{\bot},\tau\right)  =\left(  \sqrt{2}%
\zeta_{2}-f_{21}q_{1}-f_{22}q_{2}\right)  \Phi_{\zeta_{\bot}}\left(
\mathbf{q}_{\bot},\tau\right)  . \label{A.3}%
\end{equation}
Comparing the Eqs. (\ref{A.3}) and (\ref{A.2}) we have only one possibility to
make the following identifications%
\begin{equation}
w\left(  x,y\right)  =\Phi_{\zeta_{\bot}}\left(  \mathbf{q}_{\bot}%
,\tau\right)  ,\text{ \ }a=g_{21},\text{ \ }b=g_{22},\text{ \ }q_{1}=x,\text{
\ }q_{2}=y. \label{A.4}%
\end{equation}
But, for the functions $f\left(  x\right)  \longrightarrow f\left(
q_{1}\right)  $ and $g\left(  y\right)  \longrightarrow g\left(  q_{2}\right)
$ there is more than one possibility, for example,%
\begin{align}
I.\text{ \ }f\left(  q_{1}\right)   &  =-f_{21}q_{1},\text{ \ }g\left(
q_{2}\right)  =\sqrt{2}\zeta_{2}-f_{22}q_{2},\nonumber\\
II.\text{ \ }f\left(  q_{1}\right)   &  =\sqrt{2}\zeta_{2}-f_{21}q_{1},\text{
\ }g\left(  q_{2}\right)  =-f_{22}q_{2},\nonumber\\
III.\text{ \ }f\left(  q_{1}\right)   &  =\frac{\zeta_{2}}{\sqrt{2}}%
-f_{21}q_{1},\text{ \ }g\left(  q_{2}\right)  =\frac{\zeta_{2}}{\sqrt{2}%
}-f_{22}q_{2}. \label{A.5}%
\end{align}
Any of the possibilities listed above leads to the same solution of the Eqs.
(\ref{c.1a}). Let us consider the situation $III$. Thus, the general solution
of the Eq. (\ref{A.3}) takes the form%
\begin{align}
&  \Phi_{\zeta_{\bot}}\left(  \mathbf{q}_{\bot},\tau\right)  =\Upsilon\left(
u\right)  \exp\left(  \frac{\zeta_{2}q_{1}}{\sqrt{2}g_{21}}+\frac{\zeta
_{2}q_{2}}{\sqrt{2}g_{22}}-\frac{f_{21}}{g_{21}}\frac{q_{1}^{2}}{2}%
-\frac{f_{22}}{g_{22}}\frac{q_{2}^{2}}{2}\right)  ,\nonumber\\
&  u=g_{22}q_{1}-g_{21}q_{2}. \label{A.6}%
\end{align}
At the same time, the solution (\ref{A.6}) must be satisfy the Eq.
(\ref{c.1a}) with $\beta=1$,
\begin{equation}
\left(  f_{11}q_{1}+f_{12}q_{2}+g_{11}\partial_{q_{1}}+g_{12}\partial_{q_{2}%
}\right)  \Phi_{\zeta_{\bot}}\left(  \mathbf{q}_{\bot},\tau\right)  =\sqrt
{2}\zeta_{1}\Phi_{\zeta_{\bot}}\left(  \mathbf{q}_{\bot},\tau\right)  .
\label{A.7}%
\end{equation}
Using the relation $f_{11}g_{21}-f_{21}g_{11}=f_{22}g_{12}-f_{12}g_{22}$ and
the transformations%
\begin{equation}
\partial_{q_{1}}=g_{22}\partial_{u}+g_{22}^{\ast}\partial_{u^{\ast}},\text{
\ }\partial_{q_{2}}=-g_{21}\partial_{u}-g_{21}^{\ast}\partial_{u^{\ast}}.
\label{A.8}%
\end{equation}
We find that (\ref{A.6}) is a solution of the Eq. (\ref{A.7}) if the function
$\Upsilon\left(  u\right)  $ satisfies the following equation
\begin{align}
&  \left(  \partial_{u}+\frac{F}{g_{21}g_{22}}u+\frac{F_{0}}{\sqrt{2}%
g_{21}g_{22}}\right)  \Upsilon\left(  u\right)  =0,\nonumber\\
&  F=\frac{f_{11}g_{21}-f_{21}g_{11}}{g_{11}g_{22}-g_{12}g_{21}},\text{
\ }F_{0}=\frac{\left(  g_{11}g_{22}+g_{12}g_{21}\right)  \zeta_{2}%
-2g_{21}g_{22}\zeta_{1}}{g_{11}g_{22}-g_{12}g_{21}}, \label{A.9}%
\end{align}
whose general solution reads%
\begin{equation}
\Upsilon\left(  u\right)  =\exp\left(  -\frac{F}{g_{21}g_{22}}\frac{u^{2}}%
{2}-\frac{F_{0}}{\sqrt{2}g_{21}g_{22}}u+i\phi_{\bot}\right)  , \label{A.10}%
\end{equation}
where $\phi_{\bot}\left(  \tau\right)  $ is an arbitrary time-dependent function.

\section{Calculating the states $\Phi_{n}\left(  \mathbf{q},\tau\right)  $}

The number states $\Phi_{n}\left(  \mathbf{q},\tau\right)  =\left\langle
\mathbf{q}|\mathbf{n},\tau\right\rangle $ we find from (\ref{c.21}) as follows%
\begin{equation}
\Phi_{n}\left(  \mathbf{q},\tau\right)  =\frac{\left(  \hat{A}_{1}^{\dagger
}\right)  ^{n_{1}}\left(  \hat{A}_{2}^{\dagger}\right)  ^{n_{2}}\left(
\hat{A}_{3}^{\dagger}\right)  ^{n_{3}}}{\sqrt{n_{1}!n_{2}!n_{3}!}}\Phi
_{0}\left(  \mathbf{q},\tau\right)  , \label{B.1}%
\end{equation}
where the vacuum state $\Phi_{0}\left(  \mathbf{q},\tau\right)  $ is given in
(\ref{c.28}).

The creation operators $\hat{A}_{j}^{\dagger}\left(  \tau\right)  $ we can
write as follows
\begin{align}
\hat{A}_{1}^{\dagger}  &  =\frac{f_{11}^{\ast}\left(  q_{1}-iq_{2}\right)
-g_{11}^{\ast}\left(  \partial_{q_{1}}-i\partial_{q_{2}}\right)  }{\sqrt{2}%
}=-\frac{g_{11}^{\ast}\Phi_{\bot}^{-1}\left(  \partial_{q_{1}}-i\partial
_{q_{2}}\right)  \Phi_{\bot}}{\sqrt{2}},\nonumber\\
\hat{A}_{2}^{\dagger}  &  =\frac{f_{22}^{\ast}\left(  q_{1}+iq_{2}\right)
-g_{22}^{\ast}\left(  \partial_{q_{1}}+i\partial_{q_{2}}\right)  }{i\sqrt{2}%
}=-\frac{g_{22}^{\ast}\Phi_{\bot}^{-1}\left(  \partial_{q_{1}}+i\partial
_{q_{2}}\right)  \Phi_{\bot}}{i\sqrt{2}},\nonumber\\
\hat{A}_{3}^{\dagger}  &  =\frac{f_{0}^{\ast}q_{3}-g_{3}^{\ast}\partial
_{q_{3}}}{\sqrt{2}}+\varphi_{3}^{\ast}\left(  \tau\right)  =-\frac{g_{3}%
^{\ast}\Phi_{3}^{-1}\partial_{q_{3}}\Phi_{3}}{\sqrt{2}}, \label{B.2}%
\end{align}
wherein the functions $\Phi_{\bot}$ and $\Phi_{3}$ are given by%
\begin{equation}
\Phi_{\bot}=\exp\left(  -\frac{f_{11}^{\ast}}{g_{11}^{\ast}}\frac{q_{1}%
^{2}+q_{2}^{2}}{2}\right)  ,\text{ \ }\Phi_{3}=\exp\left(  -\frac{f_{0}^{\ast
}}{g_{3}^{\ast}}\frac{q_{3}^{2}}{2}-\frac{\sqrt{2}\varphi_{3}^{\ast}}%
{g_{3}^{\ast}}q_{3}\right)  . \label{B.3}%
\end{equation}
Using the relations, see \cite{Ism2014,Ism2015,Yua2015,Gra2007}, below%
\begin{align}
&  H_{n,m}\left(  Z,Z^{\ast}\right)  =\left(  -1\right)  ^{n+m}e^{ZZ^{\ast}%
}\partial_{Z^{\ast}}^{n}\partial_{Z}^{m}e^{-ZZ^{\ast}},\text{ \ }Z\in%
%TCIMACRO{\U{2102} }%
%BeginExpansion
\mathbb{C}
%EndExpansion
,\nonumber\\
&  H_{n}\left(  x\right)  =\left(  -1\right)  ^{n}e^{x^{2}}d_{x}^{n}e^{-x^{2}%
}, \label{B.4}%
\end{align}
where $H_{n}\left(  x\right)  $ are the Hermite polynomials and $H_{n,m}%
\left(  Z,Z^{\ast}\right)  $ are the Hermite polynomials of two variables, we
find that%

\begin{align}
&  \left(  \hat{A}_{1}^{\dagger}\right)  ^{n_{1}}\left(  \hat{A}_{2}^{\dagger
}\right)  ^{n_{2}}\Phi_{\bot}^{\ast}=\left(  \frac{g_{11}^{\ast}}{g_{11}%
}\right)  ^{n_{1}}\left(  \frac{g_{22}^{\ast}}{ig_{11}^{\ast}}\right)
^{n_{2}}H_{n_{1},n_{2}}\left(  \frac{q_{1}-iq_{2}}{\sqrt{2}g_{11}^{\ast}%
},\frac{q_{1}+iq_{2}}{\sqrt{2}g_{11}}\right)  \Phi_{\bot}^{\ast},\nonumber\\
&  \left(  \hat{A}_{3}^{\dagger}\right)  ^{n_{3}}\Phi_{3}^{\ast}=\left(
\frac{g_{3}^{\ast}}{2g_{3}}\right)  ^{\frac{n_{3}}{2}}H_{n_{3}}\left(
\frac{q_{3}+\sqrt{2}\operatorname{Re}\left(  \varphi_{3}g_{3}^{\ast}\right)
}{\left\vert g_{3}\right\vert }\right)  \Phi_{3}^{\ast}. \label{B.5}%
\end{align}
Then, the sum (\ref{B.1}) takes the form%
\begin{align}
\Phi_{n}\left(  \mathbf{q},\tau\right)   &  =\frac{1}{\sqrt{n_{1}!n_{2}%
!n_{3}!}}\left(  \frac{g_{3}^{\ast}}{2g_{3}}\right)  ^{\frac{n_{3}}{2}%
}H_{n_{3}}\left(  \frac{q_{3}+\sqrt{2}\operatorname{Re}\left(  \varphi
_{3}g_{3}^{\ast}\right)  }{\left\vert g_{3}\right\vert }\right) \nonumber\\
&  \times\left(  \frac{g_{11}^{\ast}}{g_{11}}\right)  ^{n_{1}}\left(
\frac{g_{22}^{\ast}}{ig_{11}^{\ast}}\right)  ^{n_{2}}H_{n_{1},n_{2}}\left(
\frac{q_{1}-iq_{2}}{\sqrt{2}g_{11}^{\ast}},\frac{q_{1}+iq_{2}}{\sqrt{2}g_{11}%
}\right)  \Phi_{0}\left(  \mathbf{q},\tau\right)  . \label{B.6}%
\end{align}


\begin{thebibliography}{99}                                                                                               %


\bibitem {Kla1968}J. R. Klauder and E. C. Sudarshan \emph{Fundamentals of
Quantum Optics} (Benjamin, 1968).

\bibitem {Gil1973}R. Gilmore 1973 \emph{Geometry of symmetrized states} Ann.
Phys\textit{. }(NY), \textbf{74}, 391.

\bibitem {Kla1985}J. R. Klauder and B. S. Skagerstam \emph{Coherent States,
Applications in Physics and Mathematical Physics} (World Scientific,
Singapore, 1985).

\bibitem {Per1986}A. M. Perelomov \emph{Generalized Coherent States and Their
Applications} (Springer-Verlag, 1986).

\bibitem {Ali2000}S. T. Ali, J. P. Antoine and J. P. Gazeau \emph{Coherent
states, wavelets and their generalizations} (Springer-Verlag, New York,
Berlin, Heidelberg, 2000).

\bibitem {Nie2000}M. A. Nielsen and I. L. Chuang \emph{Quantum computation and
quantum information} (Cambridge University Press, Cambridge, England, 2000).

\bibitem {Gaz2009}J. P. Gazeau \emph{Coherent States in Quantum Physics}
(Wiley-VCH, Berlin, 2009).

\bibitem {Mal1979}I. A. Malkin and V. I. Man'ko \emph{Dynamical symmetries and
coherent states of quantum systems} (Nauka, Moscow, 1979).

\bibitem {Dod1987}V. V. Dodonov and V. I. Man'ko{\small \ }\emph{Invariants
and correlated states of nonstationary quantum systems}. In:
\textbf{Invariants and the Evolution of Nonstationary Quantum Systems}.
Proceedings of Lebedev Physics Institute \textbf{183},\/ M. A. Markov, ed.
(Nauka, Moscow 1987) 71-181, [translated by Nova Science, Commack, New York,
1989, 103].

\bibitem {Bag2015}V. G. Bagrov, D. M. Gitman and A. S. Pereira 2015
\emph{Coherent states of systems with quadratic hamiltonians} Braz. J. Phys.
\textbf{45}, 369.

\bibitem {Bag2013}V. G. Bagrov, D. M. Gitman, E. S. Macedo and A. S. Pereira
2013 \emph{Coherent states of inverse oscillators and related problems} J.
Phys. A: Math. Theor. \textbf{46}, 325305.

\bibitem {Bag2014}V. G. Bagrov, D. M. Gitman and A. S. Pereira 2014
\emph{Coherent and semiclassical states of a free particle} Phys. Usp.
\textbf{57} (9), 891.

\bibitem {Sch1930}E. Schr\"{o}dinger 1930 \emph{Zum Heisenbergschen
unsch\"{a}rfeprinzip Sitzungsberichte} Preus. Akad. Wiss., Phys. Math. Kl.
\textbf{19},\textbf{ }296.

\bibitem {Rob1930}H. P. Robertson 1930 \emph{A general formulation of the
uncertainty principle and its classical interpretation} Phys. Rev \textbf{35}, 667.

\bibitem {Lou1987}R. Loudon and P. L. Knight 1987 \emph{Squeezed light} J.
Mod. Opt. \textbf{34}, 709.

\bibitem {Dar1927}C. G. Darwin 1927 \emph{Free motion in the wave mechanics}
Proc. R. Soc. Lond. A \textbf{117}, 258.

\bibitem {Lan1930}L. Landau 1930 \emph{Diamagnetismus der Metalle} Zeitschrift
f\"{u}r Physik. Bd. 64.

\bibitem {Joh1949}M. H. Johnson and B. A. Lippmann 1949 \emph{Motion in a
constant magnetic field} Phys. Rev. \textbf{76}, 828.

\bibitem {Sey1965}P. W. Seymour, R. B. Leipnik and A. F. Nicholson 1965
\emph{Charged particle motion in a time-dependent axially symmetric magnetic
field} Aust. J. Phys. \textbf{18}, 553.

\bibitem {Ber1982}J. Bergou and F. Ehlotzky 1982 \emph{Charged-particle
scattering in a magnetic and a laser field and nonlinear bremsstrahlung} Phys.
Rev. A \textbf{26}, 470.

\bibitem {Mal1968}I. A. Malkin and V. I. Man'ko 1968 \emph{Coherent states of
a charged particle in a magnetic field} Zh. Eksp. Teor. Fiz. \textbf{55}, 1014.

\bibitem {Mal1969}I. A. Malkin, V. I. Man'ko and D. A. Trifonov 1969
\emph{Invariants and the evolution of coherent states for a charged particle
in a time-dependent magnetic field} Phys. Lett. A \textbf{30}, 414.

\bibitem {Dod1971}V. V. Dodonov, I. A. Malkin and V. I. Man'ko 1971
\emph{Coherent states of a charged particle in a time-dependent uniform
electromagnetic field of a plane current} physica \textbf{59}, 241.

\bibitem {Var1984}S. Varr\'{o} 1984 \emph{Coherent states of an electron in a
homogeneous constant magnetic field and the zero magnetic field limit} J.
Phys. A: Math. Gen. \textbf{17}, 1631.

\bibitem {Rem2005}K. Kowalski and J. Rembielinski 2005 \emph{Coherent states
of a charged particle in a uniform magnetic field} J. Phys. A: Math. Gen.
\textbf{38}, 8247.

\bibitem {Set2009}M. R. Setare and A. Fallahpour 2009 \emph{Generalized
coherent states for charged particle in uniform and variable magnetic field}
ACTA Phys. Pol. B \textbf{40}, 217.

\bibitem {Alf1965}V. De Alfaro and T. Regge \emph{Potential scattering}
(North-Holland Publish., Amsterdam 1965).

\bibitem {Hel1975}E. J. Heller 1975 \emph{Time-dependent approach to
semiclassical dynamics} J. Chem. Phys. \textbf{62}, 1544.

\bibitem {Hel1981}E. J. Heller 1981 \emph{Frozen gaussians: A very simple
semiclassical approximation }J. Chem. Phys. \textbf{75}, 2923.

\bibitem {Lit1986}R. G. Littlejohn 1986 \emph{The semiclassical evolution of
wave packets} Phys. Rep. \textbf{138}, 193.

\bibitem {Gro1998}F. Grossmann 1998 \emph{Semiclassical coherent-state path
integrals for scattering} Phys. Rev. A \textbf{57}, 3256.

\bibitem {Mil2001}W. H. Miller 2001 \emph{The semiclassical initial value
representation: A potentially practical way for adding quantum effects to
classical molecular dynamics simulations} J. Phys. Chem. A \textbf{105}, 2942.

\bibitem {Bar2001}M. Baranger, M. A. M. de Aguiar, F. Keck, H. J. Korsch and
B. Schellhaa\ss \ 2001 \emph{Semiclassical approximations in phase space with
coherent states} J. Phys. A: Math. Gen. \textbf{34}, 7227.

\bibitem {Bag2011}V. G. Bagrov, S. P. Gavrilov, D. M. Gitman and D. P. Meira
Filho 2011 \emph{Coherent and semiclassical states in magnetic field in the
presence of the Aharonov-Bohm solenoid} J. Phys. A: Math. Theor. \textbf{44}, 055301.

\bibitem {Ohs2013}T. Ohsawa and M. Leok 2013 \emph{Symplectic semiclassical
wave packet dynamics} J. Phys. A: Math. Theor. \textbf{46}, 405201.

\bibitem {Ozo2016}A. M. Ozorio de Almeida and O. Brodier 2016
\emph{Semiclassical evolution of correlations between observables} J. Phys. A:
Math. Theor. \textbf{49}, 185302.

\bibitem {Pol2002}A. D. Polyanin, V. F. Zaitsev and A. Moussiaux\emph{
Handbook of First Order Partial Differential Equations} (Taylor \& Francis,
London 2002).

\bibitem {Ism2014}M. E. H. Ismail and P. Simeonov 2014 \emph{Complex hermite
polynomials: their combinatorics and integral operators} Proc. Amer. Math.
Soc. \textbf{143}, 1397.

\bibitem {Ism2015}M. E. H. Ismail 2015 \emph{Analytic properties of complex
hermite polynomials} Trans. Amer. Math. Soc. Article electronically published.

\bibitem {Yua2015}Y. Xu 2015 \emph{Complex versus real orthogonal polynomials
of two variables} Integral Transforms and Special Functions, \textbf{26}:2, 134.

\bibitem {Gra2007}I. S. Gradshteyn and I. M. Ryzhik 2007 \emph{Table of
integrals, series, and products} (Elsevier Inc, Oxford).
\end{thebibliography}
\end{document}